\documentclass[11pt,a4paper]{article}

\pdfoutput=1

\usepackage[font=small,labelfont=bf]{caption}
\usepackage{subcaption}

\usepackage[top=80pt,bottom=85pt,left=85pt,right=85pt]{geometry}
\usepackage{amsmath,amssymb,graphicx,float,color,tikz,cite}
\usepackage[debug,pageanchor=false]{hyperref}
\definecolor{link}{rgb}{.8,.15,.1}
\hypersetup{colorlinks=true,linkcolor=link,citecolor=link,urlcolor=link,linktocpage}

\makeatletter
\@addtoreset{equation}{section}
\makeatother

\newcommand\blfootnote[1]{
  \begingroup
  \renewcommand\thefootnote{}\footnote{#1}%
  \addtocounter{footnote}{-1}%
  \endgroup
}

\begin{document}


\begin{titlepage}
	
	\begin{flushright}
     
        \end{flushright}

	\begin{center}

	\vskip .5in  
	\noindent

	{\Large \bf{
    Entanglement surfaces for rotating cylindrical black holes}}

	\vspace{1.5cm}

Fabio Billiato$^{1,2}$ \quad \& \quad Alessandra Gnecchi$^{2}$ \blfootnote{\tt \\ fabio.billiato@pd.infn.it\\ alessandra.gnecchi@pd.infn.it} \\

	{ 

	\center{${}^{1}$\textit{ Dipartimento di Fisica e Astronomia “Galileo Galilei”, Università di Padova,\\ Via Marzolo, 8, 35131 Padova, Italy}}\\
    \center{${}^2$\textit{ INFN, Sezione di Padova, Via Marzolo, 8, 35131 Padova, Italy}}

	}

	\vskip 2cm 
	     	{\bf Abstract }
	
	\vskip .1in
	\end{center}

	\noindent

We construct entanglement surfaces for rotating cylindrical black holes in a double holographic setup, extending previous results to the case of stationary backgrounds. We analyze both the 5d braneworld construction as well as the embedding in 10d type IIB string theory. We couple the rotating cylindrical black hole to a non-gravitating bath, and  study island and Hartman-Maldacena surfaces. Properties of island surfaces are characterized by three regimes, bounded by two critical parameters. In addition to the critical value known for the static case, we find that a new one emerges, related to the extremal limit of the rotating black hole. This behaviour is present both for the bottom-up as well as the top-down models, for which we find qualitative agreement.

	\noindent

	\vfill
	\eject

	\end{titlepage}

\tableofcontents

\section{Introduction}

One of the main obstacles in studying Hawking evaporation in the context of AdS/CFT holography is the property that large black holes in Anti de Sitter are in thermal equilibrium with the outside environment, since the boundary of spacetime acts as a reflecting box 
\cite{Hawking:1982dh}. In the absence of a natural outgoing flux at infinity, one has to modify the setup by inducing a leaking of radiation across the boundary \cite{Rocha:2008fe,Mathur:2014dia}, for example by coupling the black hole with external radiation, acting like a bath where the black hole can radiate into. 
The dual interpretation of this setup is a CFT which is coupled to an external system. Such construction has led to important developments in the calculation of the entanglement entropy of black hole radiation, showing that a correct accounting of quantum effects in the definition of entangling surfaces can reproduce the behavior of the Page curve of black hole radiation, a necessary behavior for the evaporation process to preserve unitarity \cite{Penington:2019npb,Almheiri:2019psf}. These findings rely on the derivation of a new kind of extremal surfaces \cite{Engelhardt:2014gca,Faulkner:2013ana}, which results in surfaces disconnected from the radiation region, called \emph{islands} \cite{Almheiri:2019hni,Almheiri:2020cfm}.  It has recently been argued that such entanglement islands allow for a consistent realization of holography in presence of global symmetries in the bulk EFT \cite{Geng:2025gns}.

The study of entanglement surfaces  
is thus crucial to our understanding of black hole evaporation, but their explicit construction is not always at reach. 
In the first works \cite{Almheiri:2019hni}, the coupling of the black hole with an external CFT has been realized by considering a black hole on an Anti de Sitter brane, embedded on a higher dimensional AdS spacetime. This construction can actually be obtained via an end-of-the-world (ETW) brane that intersects the boundary of the higher dimensional bulk spacetime at an angle $\mu_0$, 
realizing a so-called \emph{double holographic construction}  \cite{Karch:2000ct,Karch:2000gx,Chen:2020uac,Chen:2020hmv,Geng:2020fxl,Rozali:2019day}. Within this framework, island surfaces are geometries that connect the black hole on the brane and the boundary radiation region via the bulk spacetime, and can be explicitly constructed, following the HRT prescription, \cite{Almheiri:2019psy}. Moreover, thanks to the C-metric construction, it is possible to access quantum properties of black holes in 3d Anti de Sitter on a brane \cite{Emparan:2020znc,Emparan:2021yon,Bhattacharya:2025tdn}.

The bottom-up double holographic construction consists of localizing gravity in a 4d-5d braneworld à la Karch-Randall \cite{Karch:2000ct}, while string theory top-down realizations consist of an AdS$_4$ region, containing the black hole, connected to an asymptotic AdS$_5$ region  in branes construction dual to conformal defects and interfaces \cite{DHoker:2007zhm,DHoker:2007hhe,Aharony:2011yc,Assel:2011xz,Uhlemann:2021nhu,Demulder:2022aij} (for recent study of localized gravity and ETW branes in string theory we refer to \cite{Raamsdonk:2020tin,VanRaamsdonk:2021duo,Karch:2022rvr,DeLuca:2023kjj,Huertas:2023syg,Chaney:2024bgx,Anastasi:2025puv}). 

The advantage of relying on a top-down model is control over the holographic description.
However, because of the lack of a richer truncation of the 10d theory to 4 dimensions, which would allow us to consider general fields in the EFT, e.g. Maxwell and scalar fields, this setup limits the possible configurations one can study\footnote{Advances in this direction have appeared very recently \cite{Rovere:2025jks}.}. In fact, other than the case of a black brane \cite{Uhlemann:2021nhu,Demulder:2022aij}, any other black hole configuration which admits the same uplift in the 10d metric must solve 4d Einstein equations with a negative cosmological constant, thus excluding the possibility of Reissner-Nordstrom AdS black holes.
One possibility is to consider Kerr-AdS black holes, but in this case one should derive extremal surfaces which, due to the angular fibration in the metric, require us to solve complicated integro-differential equations in the braneworld setup. In fact, island surfaces for Kerr and rotating BTZ black holes have been found using 2d constructions, outside of the braneworld setup \cite{Nian:2019buz,Nian:2023xmr,Yu:2021rfg,Wang:2024itz,Yu:2025euq}. Thermodynamic and holographic properties of quantum Kerr-AdS black holes in 3d have been analyzed in braneworld setups in \cite{Bhattacharya:2025tdn}, but extremal surfaces have not been constructed. The possibility we consider is to modify the boundary of AdS, written in Poincaré coordinates, by compactifying one direction to $S^1$, thus allowing us to analyze  a cylindrical black hole \cite{Lemos:1994xp}. This solution is a generalization of the  previously studied 4d static black brane, with one spatial direction compactified, that admits an additional parameter associated with the rotation. 

In this work we study the behavior of extremal surfaces in presence of rotation. Interestingly, adding rotation allows us to consider an extremal limit of the cylindrical black hole, which introduces a new critical value of the system's parameters. We find three regimes in the islands and entanglement phase space, bounded by our new critical parameter and by the previously found one \cite{Chen:2020hmv, Geng:2020fxl, Uhlemann:2021nhu}. We derive this both in a bottom-up braneworld construction, by embedding the brane where the black hole sits in a 5d AdS bulk, as well as on a top-down construction engineered in 10d type IIB. The extremal limit has a topological singularity which we treat as a good singularity, as it is the limit of a family of near-extremal, regular, metrics.

Two caveats are worth to be mentioned in this setup. First of all, notice that in this construction, the boundary of the bulk spacetime contains a black hole, and the dual theory is a CFT on a black hole background. This means that the CFT fields are coupled to non-dynamical gravity and are in equilibrium with an eternal black hole \cite{Hubeny:2009ru,Hubeny:2009rc,Marolf:2013ioa}. This defines the state whose entanglement entropy is captured by the extremal surfaces in the braneworld setup and of which we study the entanglement phase structure. This also provides an \emph{intermediate picture}, peculiar of doubly holographic models, where the BCFT is realized by dualizing the defect dof in the gravity description, for which the precise dictionary has been studied recently  in\cite{Karch:2022rvr,He:2025due}. It is in this picture that the black hole plus bath system is given.
Finally, notice that transparent boundary condition in AdS modifies the conservation of the bulk stress tensor, inducing a mass for the bulk graviton \cite{{Porrati:2003sa,Porrati:2001gx,Porrati:2002dt,Aharony:2006hz}}. It has been shown how the fact that the gravitational EFT actually contains a massive graviton is crucial to reproducing a Page curve \cite{Geng:2021hlu,Geng:2023qwm,Geng:2025byh}.
These models, however, allow one to study the entanglement structure of the black hole in a tractable and systematic way. In particular, they will allow us to explore the effect of rotation on extremal surfaces in braneworld/double-holographic setups, extending previous literature to the case of non-static black holes. 

The paper is organized as follows. We review the 4d boosted black string giving rise to the cylindrical black hole at the end of this section. We embed the solution in 5d bulk in section \ref{Sez:5dBraneworldModel}, where we numerically find extremal surfaces and compare their areas as function of the rotation and  brane angle. We identify three regimes, bounded by two critical values of the brane angle, one of which is new. We discuss how islands and the entanglement structure change in each regime. In section \ref{Sez:10dModel} the cylindrical black hole is embedded in a type IIB construction. In the same section we show how the three regimes are present also in the top-down setup. We end the work with concluding remarks.

\subsection{Rotating cylindrical black holes}
A solution generating technique, consisting in the boost of a planar AdS$_4$ black hole, was used in \cite{Lemos:1994xp,Lemos:1995cm} to produce a rotating black hole with cylindrical symmetry. The starting point is the planar black hole with one compact direction
\begin{equation}
    \frac{ds^2}{\ell_4^2} = \frac{dr^2}{b(r)}+e^{2r}\left(-b(r)\,dt^2 + \frac{\ell_\phi^2}{\ell_4^2}\,d\phi^2+dz^2\right)\,,  \qquad \phi \sim \phi+2\pi\,,
    \label{eq:CylindricalStaticBH}
\end{equation}
and blackening function
\begin{equation}
    b(r) = 1-\,e^{-3(r-r_h)}\,.
\end{equation}
This metric has an event horizon at $r=r_h$ with $\mathbb{R}^{1,1}\times S^1$ topology, which is also the topology of the asymptotic boundary at $r\to+\infty$, and is a solution of pure 4d Einstein gravity with negative cosmological constant $\Lambda_4 = -6/\ell_4^2$. A spacelike curvature singularity is encountered as $r\to-\infty$.

Compared to the planar case, the solution is characterized by an additional scale, given by the radius $\ell_\phi$ of the $S^1$ factor. This allows to give a physically meaningful definition of temperature in the cylindrical solution, as opposed to the planar case\footnote{From the point of view of the dual CFT, in the planar $\mathbb{R}^2$ case, the only dimensionful parameter is the would-be temperature $T$ and conformal invariance implies that the CFT observables do not depend on it \cite{Witten:1998zw}. In the cylindrical case instead, there are two dimensionful quantities $T,\ell_\phi$ and CFT observables can depend on their dimensionless combination $\ell_\phi\,T$ \cite{Marolf:2013ioa}. This is reminiscent of the spherical case. From the gravity side, the would-be temperature is given by $T \sim e^{r_h}$. In the planar case we can freely shift $r_h$ by a coordinate re-definition: $r\to r +\log\delta $ and $\phi,z \to \delta ^{-1}\,\phi\,,\,\delta ^{-1} \,z$. In the cylindrical case, this redefinition also rescales $\ell_\phi$ in such a way that $\ell_\phi\,e^{r_h}$ is left invariant. \label{ftn:PlanarVsCylBHTemp}} \cite{Witten:1998zw}, which is given by the would-be temperature of the planar solution: $T=\frac{3}{4\pi\ell_4}e^{r_h}$ considered at a fixed value of $\ell_\phi$. 

The spatial $S^1$ factor allows the solution to support rotation, which can be introduced by performing a “boost" along $\phi$  
\cite{Lemos:1994xp}
\begin{equation}
    \begin{cases}
        t = \lambda\, \left( t' - \,\Omega \frac{\ell_\phi'}{\ell_4}\,\phi' \right)\\
        \phi = \lambda\frac{\,\ell_4}{\ell_\phi} \,\left(\frac{\ell_\phi'}{\ell_4}\phi'-\Omega\,t'\right)
    \end{cases}\, , 
    \label{eq:ChangeOfCoordToRotatingBH}
\end{equation}
where $\lambda^{-1} = \sqrt{1-\Omega^2}$, and $|\Omega| \in[0,1)$ is related to the black hole angular velocity.

In the planar case, Eq. \eqref{eq:ChangeOfCoordToRotatingBH} is a genuine boost and generates a metric equivalent to \eqref{eq:CylindricalStaticBH}. The story is different in the cylindrical case, where \eqref{eq:ChangeOfCoordToRotatingBH} is only locally but not globally defined as a coordinate change. This is obvious considering that periodic and non-periodic coordinates are mixed together. Consequently, Eq.  \eqref{eq:ChangeOfCoordToRotatingBH} is better understood as a “solution-generating" transformation that can be applied to \eqref{eq:CylindricalStaticBH} in order to obtain a new, locally equivalent, but \textit{globally different} spacetime \cite{Lemos:1994xp}.

\begin{figure}[]
\centering
\includegraphics[width=0.8\textwidth]{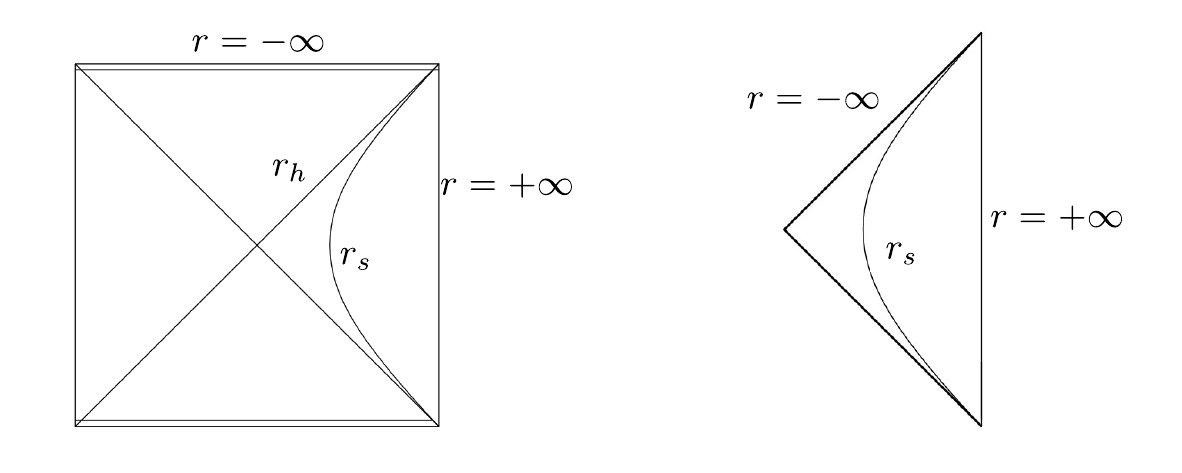}
\caption{Penrose diagram of the rotating cylindrical black hole at finite temperature (left) and at extremality (right).}
\label{fig:CylBHPenroseDiagram}
\end{figure}

The metric of the rotating cylindrical black hole obtained in this way reads (re-defining $t',\phi',\ell_\phi' \to t,\phi,\ell_\phi)$:
\begin{align}
    \frac{ds^2}{\ell_4^2} = \frac{dr^2}{b(r)} +e^{2r} \Biggl[-\frac{b(r)-\Omega^2 }{1-\Omega^2}\,dt^2 + \frac{2\Omega \bigl(b(r)-1 \bigr)}{1-\Omega^2}\,\frac{\ell_\phi\, d\phi}{\ell_4}\,dt +\frac{1-\Omega^2 b(r)}{1-\Omega^2} \frac{\ell_\phi^2 \,d\phi^2}{\ell_4^2} +dz^2\Biggr]\,.
    \label{eq:RotatingCylindricalBH}
\end{align}
The event horizon is still located at $r=r_h$ but is now generated by the killing vector $\xi = \partial_t+\frac{\Omega}{\ell_\phi}\,\partial_\phi$. Similarly to the Kerr black hole, an ergoregion appears at $r_h<r<r_s$, where $e^{3r_s}=\frac{e^{3r_h}}{1-\Omega^2} $ is such that $g_{tt}(r_s)=0$. However, although non extremal,  the metric \eqref{eq:RotatingCylindricalBH} only has one (event) horizon and lacks the internal Cauchy horizon. As discussed in \cite{Lemos:1995cm}, this implies that the rotating cylindrical black hole \eqref{eq:RotatingCylindricalBH} shares with the static solution the same global structure, shown in Fig. \ref{fig:CylBHPenroseDiagram}. This will allow us to consider, in the next section, extremal surfaces analogous to the ones of the static solution. 

The metric \eqref{eq:RotatingCylindricalBH} is left invariant under the following coordinate/parameter redefinitions (that keep $\phi$ and $\Omega$ fixed):
\begin{equation}
    r \to r+\log\delta\,, \qquad t,z\to \frac{t}{\delta},\frac{z}{\delta}\,;\qquad r_h \to r_h+\log\delta\,, \qquad\ell_\phi \to \frac{\ell_\phi}{\delta}\,, 
    \label{eq:RotatingCYlBHScalingSymmetry}
\end{equation}
meaning that only $\ell_\phi\,e^{r_h}$ and $\Omega$ are physical parameters of the solution. In this section, we are going to keep track of $\ell_\phi$ and $\ell_4$, and fix them to $\ell_\phi=\ell_4 =1$ later on.

The conserved charges, which are understood as line densities, are mass $M$ and angular momentum $J$, which, for the metric above parametrized by $r_h\,,\,\Omega$ are \cite{Lemos:1994xp,Dehghani:2002rr} ($G_N=1$)
\begin{equation}
     M = \frac{\ell_\phi\,e^{3r_h}}{8\ell_4}\,\frac{2+\Omega^2}{1-\Omega^2}\,, \qquad \qquad J=\frac{3\ell_\phi^2\,\,e^{3r_h}}{8\ell_4} \frac{\Omega}{1-\Omega^2}\,.
     \label{eq:RotatingCYlBHCharges}
\end{equation}
They satisfy the extremality bound $M\,\ell_\phi \ge J$, saturated in the extremal $\Omega \to 1$ limit, which will be  discussed shortly. 
One can also extract the temperature $T$, angular velocity $\Omega_{b}$ and entropy (line density) $S$ for the cylindrical black hole as
\begin{equation}
    T = \frac{3e^{r_h}}{4\pi\,\ell_4}\,\sqrt{1-\Omega^2}\,, \qquad \Omega_{b} = \frac{\Omega}{\ell_\phi}\,,\qquad S= \frac{\pi\ell_\phi\,e^{2r_h}}{2}\frac{1}{\sqrt{1-\Omega^2}}\,.
    \label{eq:RotatingCYlBHTempEntropy}
\end{equation}
It can be checked that the above thermodynamic quantities satisfy the first law
\begin{equation}
    dM = T\,dS+\Omega_{b}\,dJ\,.
\end{equation}
The Euclidean on-shell action (density) $I$ gives the grand-canonical thermodynamic potential $I=\beta\,G$ 
\begin{equation}
    G = M-T\,S-\Omega_{b}\,J = -\frac{8\pi^3\,\ell_\phi\,\ell_4^2}{27}\,\frac{T^3}{\bigl(1-\ell_\phi^2\,\Omega_{b}^2\bigr)^{3/2}}\,.
    \label{eq:CylBHGranCanonicalPotential}
\end{equation}
One can check that $G(T,\Omega_{b})$ is a concave function for all values of $T,\Omega_{b}$. This means that the cylindrical black hole is always locally thermodynamically stable in the grand-canonical ensemble, which is an important point for our analysis as we are interested in configurations where the black hole is in thermal equilibrium with its Hawking radiation. We also note that an Hartle-Hawking like state, describing the equilibrium state of quantum fields in the black hole background, can be defined for the rotating solution \cite{Hawking:1998kw, Hubeny:2009rc}.

Global stability of the cylindrical black hole would be undermined by a phase transition à la Hawking-Page, if another saddle point of the grand canonical partition function exists  satisfying the same boundary conditions as \eqref{eq:RotatingCylindricalBH}, which dominates in a region of parameter space. No such competing configuration has been studied so far\footnote{We just note that the static cylindrical black hole can undergo a phase transition à la Hawking-Page \cite{Surya:2001vj, Marolf:2013ioa}, with the competing solution being the Horowitz-Myers AdS soliton\cite{Horowitz:1998ha}. However, this solution only exists provided that antiperiodic boundary conditions are imposed for fermions along the spatial $S^1$ circle\cite{Surya:2001vj,Hubeny:2009rc}. If periodic boundary conditions are chosen, the AdS soliton is not an allowed solution and the static cylindrical black hole dominates the canonical ensemble, as thermal AdS$_4$ is always subdominant \cite{Marolf:2013ioa}. We may expect a similar story to hold also in the rotating solution, even if a more detailed analysis would be required. \label{ftn:HPTransitionStaticCase}}, however, notice that the results we will present are valid throughout the parameter space, in a sense that we are going to explain, so our analysis will apply even if the cylindrical black hole is only the dominant phase for a subregion of thermodynamics space.

\subsubsection*{Extremal limit}

From the black hole metric \eqref{eq:RotatingCylindricalBH} one recovers empty AdS$_4$ (with boundary $R^{1,1}\times S^1$) by taking the limit $e^{r_h} \to 0$ with fixed $\Omega$, which also sends the temperature to zero.

More interestingly, by taking at the same time  $\Omega \to1$, the extremal limit is defined by
\begin{equation}
    \Omega \to 1\,, \qquad e^{r_h} \to0\,, \qquad \frac{e^{3r_h}}{1-\Omega^2} = \frac{8\,m}{3}\,, 
    \label{eq:CylindricalBHExtremalLimit}
\end{equation}
with $m$ a fixed parameter. This limit yields a finite value for the various thermodynamic quantities 
\begin{equation}
    J = \ell_\phi\,M = \frac{\ell_\phi^2}{\ell}\,m\,, \qquad T= S = 0\,,
\end{equation}
while the metric reads
\begin{equation}
    ds^2 = \ell_4^2\,dr^2 + e^{2r}\left(-\ell_4^2\,dt^2+\ell_\phi^2\,d\phi^2+\ell_4^2\,dz^2\right) +\frac{8m}{3}\,e^{-r}\left(\ell_4\,dt-\ell_\phi\,d\phi\right)^2\,.
    \label{eq:CylindricalBHExtremalMetric}
\end{equation}
The finite temperature black hole has only one horizon; the extremal limit has the effect of pushing the horizon towards the singularity, which sits at $r=-\infty$, thus coinciding with it at $\Omega=1$. In this limit the curvature singularity disappears and the Kretschmann scalar is constant
\begin{equation}
    R_{\mu\nu\rho\sigma}R^{\mu\nu\rho\sigma} = \frac{24}{\ell_4^2}\,.
\end{equation}
The extremal metric has a “topological" singularity \cite{Lemos:1994xp}, given by the null $r=0$ line, that acts like a boundary of the spacetime beyond which geodesics cannot be extended. Consequently, the interior region of the black hole geometry disappears in this limit and one is left only with the exterior region \cite{Lemos:1995cm}, as shown by the corresponding Penrose diagram in Fig. \ref{fig:CylBHPenroseDiagram}. 
This extremal solution is supersymmetric, if embedded in $d=4$, $\mathcal{N}=2$ gauged  supergravity \cite{Lemos:2000wp}.

\section{5d braneworld model}
\label{Sez:5dBraneworldModel}

In order to discuss black hole evaporation, we couple the metric \eqref{eq:RotatingCylindricalBH} to 
a \textit{non-gravitating} bath, by embedding it in double holography \cite{Almheiri:2019hni,Almheiri:2019psy,Geng:2020qvw,Chen:2020uac,Chen:2020hmv,Geng:2020fxl,Geng:2021mic,Karch:2023ekf}. In this section we consider its 5d bottom-up realization, and we review in some detail the relation of this setup to the 5d black string metric \eqref{eq:5DBlackStringGeometry}.

\begin{figure}[]
\centering
\includegraphics[width=0.6\textwidth]{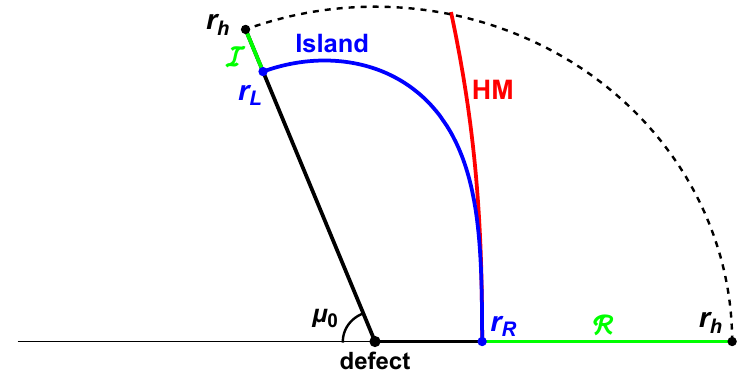}
\caption{Sketch of the black string geometry \eqref{eq:5DBlackStringGeometry}, with the two classes of bulk extremal surfaces homologous to a given bath subregion $\mathcal{R}$. Each $\mu=const.$ slice hosts the cylindrical black hole \eqref{eq:RotatingCylindricalBH}. Island surfaces end on the brane, while HM surfaces end on the horizon.}
\label{Fig:BlackStringGeometry}
\end{figure}

We consider the Karch-Randall (KR) model \cite{Randall:1999vf,Karch:2000ct}, consisting of a single subcritical \textit{thin} end-of-the-world brane embedded on a AdS$_4$ slice inside AdS$_5$. The KR brane intersects the asymptotic $\mathbb{R}^{1,3}$ boundary of AdS$_5$ at an angle $\mu_0$, removing half of it (see Fig. \ref{Fig:BlackStringGeometry}). The parameter $\mu_0$ is related to the tension of the KR brane \cite{Karch:2000ct} and represents the free parameter that characterizes the braneworld model (for a review see \cite{Chen:2020uac}).
Thanks to AdS/CFT holographic duality, this gravitational system possesses two additional equivalent descriptions\cite{Karch:2000gx,Takayanagi:2011zk,Fujita:2011fp}.
The first one is the so called \textit{intermediate picture}, which is obtained by integrating out the 5d bulk\footnote{A precise definition of this picture has been given, for the bottom-up model, in \cite{Geng:2023qwm,Geng:2025yys} and in \cite{Karch:2022rvr} for top-down realizations, resolving some inconsistencies that appears in more ``naive" definitions \cite{Omiya:2021olc,Neuenfeld:2023svs,Mori:2023swn}.} \cite{deHaro:2000wj}. 
It is described by an \textit{induced} AdS$_4$ gravity theory coupled to a holographic CFT$_4$ with a UV cutoff and communicating, via transparent boundary conditions at the boundary of AdS$_4$, to the same non-gravitating CFT$_4$ on the remaining boundary of the 5d bulk space. 
By further dualizing the 4d induced gravity theory on the brane, one gets a fully non-gravitating description, given by a BCFT consisting in the previous CFT$_4$ coupled to 3d defect degrees of freedom. The number of 3d degrees of freedom increases as $\mu_0$ decreases \cite{Chen:2020uac}.

The intermediate picture is the convenient setup to discuss black hole evaporation \cite{Almheiri:2019hni}, in particular by replacing the AdS$_4$ geometry induced on the KR brane by the (eternal) black hole geometry \eqref{eq:RotatingCylindricalBH}, which also changes the boundary topology to $\mathbb{R}^{1,2}\times S^1$. The non-gravitating CFT$_4$ serves as heath bath; it is in thermal equilibrium with the black hole and exchanges heath with it thanks to the transparent boundary conditions at the boundary of the brane. Concretely, this configuration can be realized by the following 5d black string metric (following \cite{Geng:2021mic, Karch:2023ekf})
\begin{equation}
    ds^2 = \frac{\ell_5^2}{\sin^2 \mu}\left[ds^2_{cyl} +d\mu^2\right]\,,
    \label{eq:5DBlackStringGeometry}
\end{equation}
where $ds^2_{cyl}$ is the black hole metric \eqref{eq:RotatingCylindricalBH} with $\ell_\phi = \ell_4=1$. \eqref{eq:5DBlackStringGeometry} is a solution of pure 5d gravity with negative cosmological constant. The presence of the KR, ETW brane at $\mu = \mu_0$ fixes the range of coordinates to $\mu \in[\mu_0,\pi)$, which parametrizes the 4d slicing of the bulk;  $\mu=\pi$ corresponds to the leftover half of the 5d asymptotic boundary. 
This is where the bath lives, which corresponds to a conformal theory CFT$_4$ on a non-dynamical black hole background, at the same temperature and angular velocity as the black hole induced on the KR brane, whose horizon is connected via the bulk geometry to the bath. The equilibrium state of black hole and matter fields is a Hartle-Hawking state, which is well defined also for matter fields on the rotating black hole background \eqref{eq:RotatingCylindricalBH} \cite{Hawking:1998kw,Hubeny:2009rc}. 
This geometry is shown in Fig. \ref{Fig:BlackStringGeometry}.

In the setup we've described, the Entanglement Entropy (EE) of the Hawking radiation collected by the bath is obtained by evaluating the EE associated to a given subregion $\mathcal{R}=\{r\le r_R\}$ in the bath. In the globally extendend spacetime of Fig. \ref{fig:CylBHPenroseDiagram}, where the bath CFT lives, the subregion  $\mathcal{R}$ we consider is a connected interval that intersects the bifurcation surface at $t=0$, and is symmetric with respect to this point of the Penrose diagram.
Conveniently, this quantity is captured holographically by RT/HRT \cite{Ryu:2006bv,Hubeny:2007xt} surfaces that extend in  the 5d bulk \cite{Almheiri:2019hni,Chen:2020hmv} anchoring at $\partial \mathcal{R} = \{r=r_R\}$ on the boundary. 

There are two relevant classes of extremal surfaces that compete for the black hole Page curve; Hartman-Maldacena (HM) surfaces \cite{Hartman:2013qma} (depicted in red in Fig. \ref{Fig:BlackStringGeometry}) reach the black string horizon through the 5d bulk, and in the extended spacetime of Fig. \ref{fig:CylBHPenroseDiagram} they cross into the interior region connecting to the copy of $\partial \mathcal{R} $ on the left boundary. Their area grows with the size of the interior region, and so it will grow unbounded with time evolution\footnote{The proof that this is actually the case also in the black string geometry considered here has been given in \cite{Geng:2020fxl} for the static black hole. In the rotating case, the same argument continues to hold, since the rotating solution has the same global structure as the static one.}. This is expected, since the eternal black hole and the bath exchange radiation  which is not initially entangled. As the two systems continue to exchange radiation, the entanglement between them increases over time. Such unbounded growth of the area causes inconsistencies in the accounting of black hole entropy which, by itself, can be regarded as a sign of information paradox for eternal black holes coupled to a bath, as discussed in \cite{Almheiri:2019yqk}. 

Another class of  extremal surfaces (depicted in blue in Fig. \ref{Fig:BlackStringGeometry}) consists of those that  connect the boundary region to the brane via the higher dimensional bulk, and correspond to the geometrization of the quantum extremal island surfaces $\mathcal{I}$  \cite{Engelhardt:2014gca}. These lie outside of the horizon \cite{Almheiri:2019yqk}, and in the 4d setup, they are causally disconnected to the bath subregion $\mathcal{R}$. 
In this setup their area is time independent, so they will eventually be the dominant extremal surface in the entanglement entropy computation, leading to a Page curve consistent with  unitarity. This is a consequence of the fact that the rotating black hole \eqref{eq:RotatingCylindricalBH} has the same causal structure as the static black brane.

Black string geometries described by \eqref{eq:5DBlackStringGeometry} may suffer, in principle, from Gregory-Laflamme (GL) type instabilities \cite{Gregory:1993vy,Gregory:1994bj}, leading to fragmentation along the string extended direction $\mu$. Gubser and Mitra conjectured \cite{Gubser:2000mm} that these are linked to local thermodynamic instabilities of the solution, which in the present case would correspond to local instabilities of the 4d black hole induced on the brane \cite{Chamblin:2004vr,Chen:2008vh}. In the case of the static cylindrical black string\footnote{See also \cite{Hirayama:2001bi, Marolf:2019wkz} for the static global black string case, where the 4d slices are Schwarzschild-AdS (SAdS) black holes. A GL instability appears for small enough black hole size, mirroring the local thermodynamic instability of small SAdS black holes.}, with 4d slices given by Eq. \eqref{eq:CylindricalStaticBH}, the solution is locally thermodynamically stable \cite{Dehghani:2002rr, Surya:2001vj}, and in fact GL-type unstable modes have not been found \cite{Chen:2008vh}, in accordance with the static black hole \eqref{eq:CylindricalStaticBH} always having positive specific heat. 
The rotating cylindrical black hole is also locally thermodynamically stable, as dictated by the concavity of the grand-canonical potential \eqref{eq:CylBHGranCanonicalPotential} over the whole parameter space $(T,\Omega)$. Thus, even though there is no exhaustive analysis of instabilities, in our study we assume the validity of the Gubser-Mitra conjecture, and that no unstable GL modes arise for the rotating solution, either. 

Let us finally notice that, in the intermediate picture, when the black hole  \eqref{eq:RotatingCylindricalBH} is induced on a gravitating KR brane, there could be instability towards a phase transition, corresponding to Hawking-Page-like transitions in the 5d bulk \cite{Chamblin:2004vr,Chen:2008vh}. There are no complete studies available about phase transitions for the cylindrical rotating black hole \eqref{eq:RotatingCylindricalBH} (for some details on the static case see footnote \ref{ftn:HPTransitionStaticCase}); however, the relevant properties of extremal surfaces we analyze 
are independent on $\ell_\phi$ and $T$, as we are going to show, and the interesting new features only depend on $\Omega$. Even if an instability is present in some points of the phase space $(\ell_\phi,T,\Omega)$,  our results would apply for a region where we expect the black hole phase to dominate.

\subsection{5d Extremal surfaces}

Extremal surfaces for a stationary spacetime are derived following the covariant HRT prescription \cite{Hubeny:2007xt}. This requires us to extremise the whole area functional for a generic spacelike codimension-2 hypersurface, not necessarily contained on a fixed time slice. 

In order to describe the entanglement entropy relative to a radiation region, we consider a subregion $\mathcal{R}$ on the boundary (the bath), which wraps the cylinder coordinates $\phi,z$ at fixed time $t=0$ and extended for $r\le r_R$ \cite{Geng:2021mic,Karch:2023ekf}. The relevant surfaces $\gamma$ for the extremization, according to the HRT prescription, are surfaces in the 5d bulk that reach the asymptotic boundary along $\partial \mathcal{R}$, and also wraps $\phi,z$, due to symmetries of the regions we consider.
\subsubsection{Island surfaces}
\label{Sec:IslandSurfaces}
The codimension-2 surfaces wrapping the cylinder are surfaces $\gamma$ embedded through $t=t(\mu)\,,\, r=r(\mu)$, with induced metric in 5d given by
\begin{align}
    ds_\gamma^2  = \frac{\ell_5^2\,e^{2r}}{\sin^2 \mu} \Biggl[\,dz^2 + \frac{1-\Omega^2b(r)}{1-\Omega^2}\,d\phi^2 &+ \frac{2\Omega\bigl(b(r)-1\bigr)}{1-\Omega^2} t'\,d\phi d\mu+ \notag \\
    &+ \left(e^{-2r}+\frac{e^{-2r}}{b(r)}\,r'^{\,2}-\frac{b(r)-\Omega^2}{1-\Omega^2}\,t'^{\,2}\right)d\mu^2\Biggr]\,,
\end{align}
where prime denotes derivative with respect to the $\mu$ coordinate. It is a spacelike surface provided that the following inequality holds:
\begin{equation}
    1+\frac{r'^{\,2}}{b(r)\,\sin^2 \mu}-\frac{e^{2r}\bigl(b(r)-\Omega^2\bigr)}{(1-\Omega^2)\sin^2\mu}\,t'^{\,2}>0\,,
\end{equation}
which is trivially satisfied for $t'=0$. The induced area functional on $\gamma$ reads
\begin{equation}
   A_\gamma = V_{cyl}\,\ell_5^3\int_{\mu_0}^\pi d\mu \frac{e^{2r}}{\sin^3 \mu} \sqrt{\Delta} \,,
   \label{eq:5DAreaTimeDepAreaFunctional}
\end{equation}
with:
\begin{equation}
    \Delta = \frac{1-\Omega^2b(r)}{1-\Omega^2}\left(1+\frac{r'^{\,2}}{b(r)}\right)-e^{2r}\,b(r)t'^{\,2}\,.
\end{equation}
$V_{cyl}$ arises from the integration along the cylinder $\phi,z$ and formally contains a divergent contribution from the infinite length of the $z$-coordinate. 
Since the only dependence on horizon topology in the area functional \eqref{eq:5DAreaTimeDepAreaFunctional} appears through an overall factor, 
by turning-off the rotation $\Omega\to0$ one recovers the results obtained by \cite{Geng:2021mic} for the static planar black hole.

The area functional \eqref{eq:5DAreaTimeDepAreaFunctional} inherits a scaling behavior from the black hole solution. In particular, under:
\begin{equation}
    r \to r+\log \lambda\,,\qquad t\to\frac{t}{\lambda}\,; \qquad r_h \to r_h+\log \lambda\,,\qquad\Omega \to \Omega\,,
\label{eq:PartialRescalingAreaFunctional}
\end{equation}
the area functional changes as $A_\gamma \to \lambda^2 A_\gamma$, leading to the same extremization conditions. Note that we are not rescaling the cylinder coordinates, $\ell_\phi$ has been fixed to $1$, hence \eqref{eq:PartialRescalingAreaFunctional} is not equivalent to the rescaling \eqref{eq:RotatingCYlBHScalingSymmetry}, which leaves the black hole metric invariant. Consequently, \eqref{eq:PartialRescalingAreaFunctional} generates a new black hole with rescaled charges, entropy and temperature given by:
\begin{equation} 
    M,\,J\  \ \longrightarrow  \ \ \lambda^3M\,,\lambda^3J \,,\qquad  S \ \ \longrightarrow \ \ \lambda^2 S\,, \qquad T \ \ \longrightarrow \ \ \lambda\, T\,.
\label{eq:PartialRescalingCharges}
\end{equation}
Thanks to this, we can focus on the study of properties of extremal surfaces at fixed mass $M$ and varying $\Omega$. Given the relations \eqref{eq:RotatingCYlBHCharges}-\eqref{eq:RotatingCYlBHTempEntropy} these correspond to black holes with varying $T$ and $J$.

The surfaces $\gamma$ must be found by imposing the appropriate island boundary conditions. On the asymptotic boundary $\mu = \pi$,  we must impose a Dirichlet boundary condition:
\begin{equation}
     r(\pi) = r_R\,,\qquad t(\pi) =t_0 = 0\,,
\end{equation}
ensuring that $\gamma$ reaches the boundary along $\partial \mathcal{R}$. 

Then, $\gamma$ ends on the KR brane at $\mu = \mu_0$, at a position $r(\mu_0) = r_L$ and time $t(\mu_0) = t_L$ that must be determined dynamically. This follows from the fact that the brane hosts dynamical gravity, which implies that $r_L$ and $\tau_L$ should be determined in a covariant way \cite{Geng:2020fxl}. At $\mu = \mu_0$ we then impose a Neumann boundary condition for both $r$ and $t$
\begin{equation}
    r'(\mu_0) = t'(\mu_0)=0\,,
\end{equation}
with this choice, the boundary terms arising from the variation of \eqref{eq:5DAreaTimeDepAreaFunctional} cancel.

The boundary condition $t'(\mu_0)=0$, together with the equation of motion for $t(\mu)$, imposes that $t'(\mu) =0$ along the whole surface. In fact, from the variation of the area functional, 
the equation of motion for $t(\mu)$ reads
\begin{equation}
    \delta_t A_\gamma = 0 \quad \longrightarrow  \quad \frac{d}{d\mu}\left[\frac{e^{4r}\,b(r)\,t'}{\sin^3\mu\,\sqrt{\Delta}}\right] = 0\,,
    \label{eq:5DTimeEOM}
\end{equation}
the term inside the square brackets is a constant along the surface and is equal to zero once we impose $t'(\mu_0)=0$. It follows that $t'=0$ everywhere, and the whole surface lives at a fixed time. Fixing $t'=0$, the equation of motion for $r(\mu)$ reads:
\begin{align}
    r''  -\frac{3\,r'^{\,2}\,\left(1+2\,r'\,\cot \mu\right)}{2\,b(r)} - \frac{r'\,\left(6\cot \mu+r'\right)}{2}-2b(r)+\frac{3\,\Omega^2\,\bigl(1-b(r)\bigr)\bigl(b(r)+r'^{\,2}\bigr)}{2\bigl(1-\Omega^2 b(r)\bigr)} =0\,,
    \label{eq:5DIslandODE}
\end{align}
which reduces to the one of
\cite{Geng:2021mic}, up to a change of radial coordinate, at $\Omega=0$.

This result is a direct consequence of the stationarity of the metric, the symmetry of the bath subregion which wraps the cylinder coordinates, and the Neumann boundary condition for the time coordinate. The latter is physically motivated by the presence of dynamical gravity on the brane. However, from a mathematical point of view, we may as well impose a Dirichlet boundary condition. This idea for the static case have been explored in \cite{Ghosh:2021axl}, and in our case would correspond to surfaces that reach the brane at time $t_L\ne t_R$. 
In the following, we are going to consider only the surfaces that satisfy the physically motivated Neumann boundary condition.

\subsubsection{HM surfaces}

The other class of surfaces homologous to the radiation region $\mathcal{R}$, relevant to our discussion, are Hartman-Maldacena surfaces, which, instead of anchoring on the brane at position $r_L$, reach the black hole horizon. It is convenient to discuss them in a parametrization that uses the Tortoise radial coordinate $u \in[0,+\infty)$, covering the region $r\geq r_h$, as
\begin{equation}
    du = \frac{dr}{\sqrt{b(r)}} \,, \qquad u(r) = \frac{2}{3}\,\cosh^{-1}\left(e^{\frac{3}{2}(r-r_h)}\right)\,\, \longrightarrow \,\, b(u) = \tanh^2\left(\frac{3u}{2}\right) \,.
\end{equation}
The black hole metric becomes
\begin{align}
    ds^2 = du^2 +e^{2r_h}\cosh^{4/3}\left(\frac{3u}{2}\right) \left[-\frac{b(u)-\Omega^2 }{1-\Omega^2}\,dt^2 \right. + &
   \frac{2\Omega \bigl(b(u)-1 \bigr)}{1-\Omega^2}\,d\phi\,dt +
    \nonumber\\
    &+\left.\frac{1-\Omega^2 b(u)}{1-\Omega^2}d\phi^2 +dz^2\right]\,.
    \label{eq:RotatingCylBHTortoise}
\end{align}
This parametrization accounts for the surface at the initial time, where the Einstein-Rosen bridge crosses the bifurcation surface, and has zero size. Since the time-evolution of these surfaces makes them grow in time, it will be enough to compare their areas with the one of the island surfaces, which instead have constant area, at initial time to discuss whether the HM surfaces dominate in a certain region of parameter space, before the islands take over as minima of the area functional.

HM surfaces are described by the embedding $t=t(u)$ and $\mu=\mu(u)$, with induced metric:
\begin{align}
    ds^2_\gamma = \frac{\ell_5^2}{\sin^2\mu}\Biggl[&e^{2r_h}\cosh^{4/3}\left( \frac{3u}{2}\right)\biggl(dz^2+\frac{1-\Omega^2b(u)}{1-\Omega^2}\,d\phi^2+\frac{2\Omega\bigl(b(u)-1\bigr)}{1-\Omega^2}\,t'\,d\phi\, du\biggr) + \notag \\
    &\qquad \left(1+\mu'^{\,2}-e^{2r_h}\cosh^{4/3}\left(\frac{3u}{2}\right)\frac{b(u)-\Omega^2}{1-\Omega^2}\,t'^{\,2} \right)\,du^2\Biggr]\,,
\end{align}
which yields the induced area functional
\begin{align}
    A_\gamma = \frac{\ell_5^3\,V_{cyl}\,e^{2r_h}}{\sqrt 2}\int_0^{u_R}du\,\frac{\cosh^{1/3}\left(3u/2\right)}{\sin^3\mu}\,\sqrt{\Delta_u}\,,
    \label{eq:5DTimeDepAreaFunctionalHM}
\end{align}
with:
\begin{equation}
    \Delta_u = \left(\frac{1+\Omega^2}{1-\Omega^2}+\cosh 3u\right)\bigl(1+\mu'^{\,2}\bigr) - e^{2r_h}\frac{\sinh^2 3u}{2\cosh^{2/3}\left(3u/2\right)}\,t'^{\,2}\,,
\end{equation}
where $u=0$ is the position of the event horizon and $u_R = u(r_R)$. The extremization equation for $t(u)$
\begin{equation}
    \frac{d}{du}\left[\frac{\sinh^2(3u)\,t'}{\sin^3 (\mu)\,\cosh^{1/3}(3u/2)\,\sqrt{\Delta_u}}\right] = 0\,,
\end{equation}
fixes the quantity in square brackets to a constant, which has to be zero, as can be seen by evaluating it at $u=0$. The equation is solved for $t'(u) =0$ along the whole surface.

The differential equation for $\mu(u)$ has boundary conditions  \cite{Geng:2020fxl,Uhlemann:2021nhu}
\begin{equation}
    \mu(u_R) = \pi\,,\qquad\mu'(0) = 0\,.
    \label{eq:5dHMBoudaryCondition}
\end{equation}
The condition $\mu'(0)=0$ is due to the fact that, on the initial time slice, HM surfaces cross the horizon at the bifurcation point, where they join with their mirror image in the other copy of the thermofield double. As there is no internal region jet, the surface has a reflection symmetry across $u=0$.
The resulting differential equation for HM surfaces, with $t'=0$, is 
\begin{equation}
    \mu'' +\bigl(1+\mu'^{\,2}\bigr) \left[3 \cot \mu+\frac{2-\Omega^2+2(1-\Omega^2)\cosh(3u)}{1+\Omega^2+(1-\Omega^2)\cosh(3u)}\,\tanh\left(\frac{3u}{2}\right)\,\mu'\right] = 0\,.
    \label{eq:5DHMEoM}
\end{equation}
This equation can be solved numerically and will be discussed in the next subsection. Note that the brane-angle does not enter neither in the ODE \eqref{eq:5DHMEoM} nor in the boundary condition \eqref{eq:5dHMBoudaryCondition}, meaning that HM surfaces only depend on $r_R$ and $\Omega$.

\subsection{Numerical solutions}

The differential equations  (\ref{eq:5DIslandODE}) and (\ref{eq:5DHMEoM}) can be solved numerically to determine quantum extremal surfaces. Their properties, and how they vary with rotation, will give us information on the entanglement phase structure of the system.

Exploiting the symmetry of the system under \eqref{eq:PartialRescalingAreaFunctional} we can fix one of the physical parameters in our numerical analysis. We  choose to fix $M=1/4$, with the others given by
\begin{equation}
    e^{3r_h} = 8M\frac{1-\Omega^2}{2+\Omega^2}\,, \qquad T = M^{1/3}\frac{3}{2 \pi}\frac{(1-\Omega^2)^{5/6}}{(2+\Omega^2)^{1/3}}\,, \qquad J = M\,\frac{3\,\Omega}{2+\Omega^2}\,.
    \label{eq:FixedMTDQuantities}
\end{equation}
Then, all the surfaces with the same value of $\Omega$ but a different mass density $M'$ are related to those with $M=1/4$ via \eqref{eq:PartialRescalingAreaFunctional}. In particular, the surface area transforms as 
\begin{equation}
    A(M',\Omega) = (4M')^{2/3}\,A\left(\frac{1}{4},\Omega\right)\,.
    \label{eq:AreaScaling}
\end{equation}
We could have chosen to fix either $r_h$, $T$ or $J$ equivalently, but fixing $M$ allows us to approach extremality easily. With this choice, the event horizon of the static black hole is located at $r_h=0$, and decreases monotonically with $\Omega$, with $r_h\to-\infty$ corresponding to $\Omega\to 1$. 
The relation between the two is
\begin{equation}
    \Omega^2 = \frac{2(1-e^{3r_h})}{2+e^{3r_h}}\,,
    \label{eq:OmegaRhrelation}
\end{equation}
which is invertible so either $\Omega \in[0,1)$ or $r_h\in(-\infty,0]$ can be chosen as the parameter controlling the rotation. In the following we are going to mainly use $r_h$, always keeping in mind that we work with fixed $M = 0.25$. 

As showed above, the choice of horizon topology, planar or cylindrical, does not affect the shape and properties of the extremal surfaces, but only enters as an overall factor in the area functional. It follows that in the $r_h \to 0$ limit we should recover the same results found in the static black brane. In order to highlight the main properties of these extremal surfaces and introduce the main ingredients and physical quantities that we will consider in our analysis, we first review the main features of the static solution, which we then generalize to the rotating case, referring to \cite{Geng:2021mic} for more details on the static black brane. 

 \subsubsection{Review of extremal surfaces for static black branes}
Extremal surfaces emanate from the boundary into the 5d bulk, moving towards the horizon in a monotonic way as they move trough it: $r(\mu)$ is an increasing function of $\mu$. HM surfaces reach the black hole horizon, while islands reach the brane at $r_L$ which is also their closest point to the horizon.   
For this reason, when discussing islands, we find useful to consider the quantity
\begin{equation}
\Delta r = r_L-r_R\,,
\label{eq:DeltarDefinition}
\end{equation}
which characterizes how much islands ``stretch" while moving through the 5d bulk. This quantity is invariant under \eqref{eq:PartialRescalingAreaFunctional}, and depends on $r_h$, $\mu_0$ and $r_R$. 

\begin{figure}[]
\centering
\includegraphics[width=0.45\textwidth]{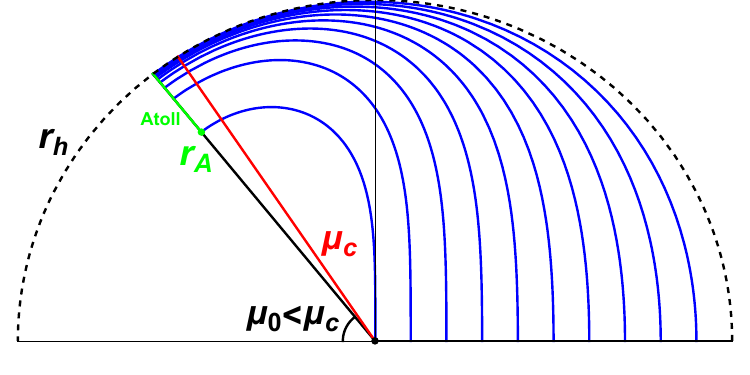}
\qquad
\includegraphics[width=0.45\textwidth]{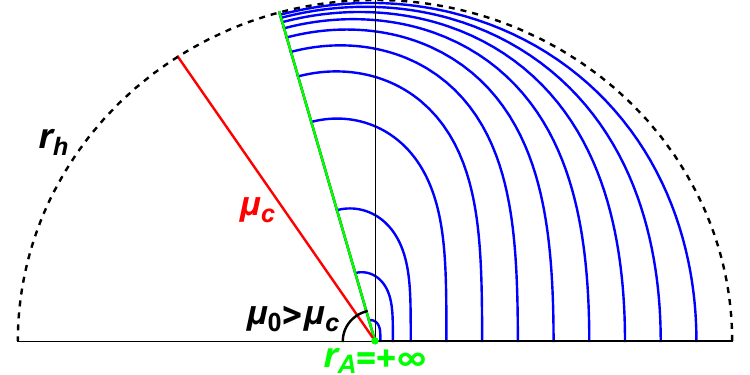}
\caption{Behavior of the island surfaces in the static black brane geometry, below (left) and above (right) the critical angle $\mu_c$. Below the critical angle, the region on the brane where islands can end, i.e. the atoll, depicted in green, does not cover the whole brane. }
\label{fig:5DAtollVsNoAtollPhases}
\end{figure}

\paragraph{Critical angle $\mu_c$:} The system shows there exist a critical (brane-)angle\cite{Chen:2020hmv, Geng:2020fxl} $\mu_c$, across which the qualitative behavior of the island surfaces display a sharp transition. At zero temperature, i.e. for empty AdS, islands only exist above\footnote{This is true for islands that do not start at the boundary exactly at the defect. In that case, in fact, at zero temperature it was found that islands that shoot into the bulk from the defect exist only at $\mu_0 = \mu_c$. This has been used to find an analytic expression for the critical angle $\mu_c$ in \cite{Chen:2020hmv, Geng:2020fxl} .}  $\mu_c$. 
At finite temperature, islands exist also below the critical angle but the behavior of the surfaces that start close to the defect, with $r_R\gg 1$, changes as compared to the $\mu_0>\mu_c$ regime, due to the formation of the \textit{atoll} \cite{Geng:2021mic, Karch:2023ekf} (See next paragraph and Fig. \ref{fig:5DAtollVsNoAtollPhases}). A diagnostic of this, is given at finite temperature by the change of behavior of $\Delta r$ at large $r_R$, which changes from being  constant above $\mu_c$ to diverging linearly with $r_R$:
\begin{equation*}
        \Delta r\, \xrightarrow[r_R\to+\infty]{}\, \Delta r^+ \quad\longleftrightarrow \quad\,r_L \sim r_R -\Delta r^+ \,, \qquad \text{for:} \quad \mu_0>\mu_c\,, 
\end{equation*}
\begin{equation}
        \Delta r \xrightarrow[r_R\to+\infty]{} r_R-r_A \qquad \longleftrightarrow \,\quad r_L \sim r_A\,, \qquad\text{for:} \quad \mu_0<\mu_c\,.
        \label{eq:IslandsDeltarBehavior}
\end{equation}
$\Delta r^+$ is a constant fixed by the system at $\mu_0$ (see Fig. \ref{Fig:5DDeltaPlusDeltaMin}), and also corresponds to the constant value of $\Delta r$ at zero temperature. The fact that, at zero temperature, $\Delta r$ does not depend on $r_R$ is a consequence of the symmetry of the corresponding differential equation under shifts of\footnote{As can be trivially seen by setting $b(r)=1$ in the ODE for islands in Eq. \eqref{eq:5DIslandODE}. Indeed, with $b(r)=1$ the 4d static/rotating black hole solution \eqref{eq:RotatingCylindricalBH} reduces to empty AdS$_4$.} $r$.

The numerical value of $\mu_c$ found in \cite{Geng:2020fxl} is
\begin{equation}
    \mu_c \sim0.98687\,,
    \label{eq:MucValue}
\end{equation}
and does not depend on the details of the induced 4d geometry on the brane, but is rather an intrinsic property of the braneworld model that appears in entanglement studies (indeed it is found to change if DPG terms are added on the KR brane \cite{Chen:2020hmv}). We expect that this ``universality" property of $\mu_c$ continues to hold also as we add rotation.

\paragraph{Atoll and critical anchor:} 
Depending on the parameters, regions on the brane appear where islands cannot end. These define the \emph{atoll}, namely the only part of the brane where islands can land on, arriving from the bulk \cite{Geng:2021mic}, see Fig. \ref{fig:5DAtollVsNoAtollPhases}. The, possibly multiple, boundaries of the atoll define the critical anchor(s) $r_A$.

For finite temperature static solutions it was found that the atoll appears below the critical angle. Its appearance is signaled by the linear divergence of $\Delta r$ at large $r_R$ as in Eq. \eqref{eq:IslandsDeltarBehavior}. The shape of the atoll, and the number of critical anchors, depend on the details of the induced 4d geometry on the brane. For the static planar black hole, there is only one critical anchor $r_A(\mu_0)$ whose value depends only on $\mu_0$, and the atoll is defined by $r_L \in(r_h,r_A)$ \cite{Geng:2021mic}. The size of the atoll is found to increase monotonically with the brane angle, reaching its maximal extension at (and above) the critical angle where it covers the whole brane, while it shrinks to the horizon as $\mu_0\to0$: 
\begin{equation}
    r_A(\mu_0\to0) = r_h\,,\qquad r_A(\mu_c) = +\infty\,.
\end{equation}

A richer structure is found for different black holes, for which $r_A$ depends on all the parameters describing it (see \cite{Karch:2023ekf} for the case of Schwarzschild-AdS$_4$). In the present case, we expect $r_A$ to depend on $\mu_0$ and $r_h$.

\paragraph{Constant entropy belt $\mathcal{C}$ and Page angle $\mu_p$:} Comparing the areas of HM and island surfaces\footnote{The area of the extremal surfaces contain two divergent contributions. One is contained in the infinite volume of the cylinder $V_{cyl}$, and one is related to the IR divergence arising as the $\mu=\pi$ boundary is approached, which is related to the usual UV divergence of entanglement entropy in QFT. We get rid of the first divergent term by considering $A_{isl/HM} = A_\gamma/V_{cyl}$. The other divergence then cancels when taking the difference $A_{isl}-A_{HM}$ upon regulating the integrals with a cutoff on $\mu$ .} at $t=0$: 
\begin{equation}\label{Area-diff}
    \Delta A(0) = A_{isl} - A_{HM}(0)\,,
\end{equation}
provides information about the entanglement phase structure of the dual BCFT state. 

If $\Delta A(0)$ is negative for a given subregion $\mathcal{R}$, i.e. for a given value of $r_R$, then the island surface dominates at all times, leading to a constant Page curve for $\mathcal{R}$. Related to this property is the concept of the \textit{constant entropy belt} which we denote as $\mathcal{C}$, first introduced in \cite{Geng:2021mic}, which can be used to summarize the entanglement phase structure of the BCFT. The constant entropy belt is defined as the region in the bath for which $\Delta A(0)$ is negative. It follows that all subregions that end inside of it, $\mathcal{\partial R}\subset\mathcal{C}$, have a constant Page curve, as opposed to subregions ending in the complement of the belt whose entanglement initially grows and saturates only at later times. The size of the belt depends on the brane-angle $\mu_0$ and is found to exist only above a certain value called the \textit{Page angle} $\mu_p$ in \cite{Geng:2021mic}. Above $\mu_p$, all bath subregions have a positive $\Delta A(0)$.

In the case of the static black brane, $\Delta A(0)$ is found to decrease monotonically with $r_R$ and $\mu_0$. This implies that, at any fixed value of $\mu_0$ above the Page angle, the constant entropy belt is a connected interval defined by:
\begin{equation}
    \mathcal{C}: \Bigl\{r_R\ge r^\star \,\Bigl| \,\, \Delta A(0)|_{r_R=r^\star} = 0\Bigr\}\,.
\end{equation}
$r^\star(\mu_0)$ characterizes the size of $\mathcal{C}$ and is found to be a monotonically decreasing function of $\mu_0$ with
\begin{equation}
r^\star(\mu_p) = +\infty\quad\longleftrightarrow\quad \Delta A(0)\Bigr|_{\substack{r=+\infty \\ \mu=\mu_p \\ \, }} = 0 \,.
\label{eq:PageAngleDefinition}
\end{equation}
Eq. \eqref{eq:PageAngleDefinition} can then be used to find the numerical value of $\mu_p$, which is found to be slightly less than $\mu_c$ in \cite{Geng:2021mic}. 

The constant entropy belt and its properties are related to the BCFT state. As we add rotation we therefore expect it to have an influence on $\mathcal{C}$.

\subsubsection{Numerical study of island surfaces}

Island surfaces for cylindrical black hole can be constructed numerically analogously to the methods used in the static black brane solution, and the shape of the surfaces remain qualitatively similar to those of the static solution. Also, adding rotation does not influence the existence of islands at finite temperature, which can always be found. 

\begin{figure}[ht!]
\centering
\includegraphics[width=0.47\textwidth]{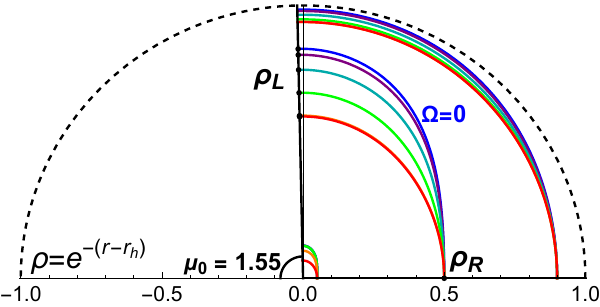}
\qquad
\includegraphics[width=0.47\textwidth]{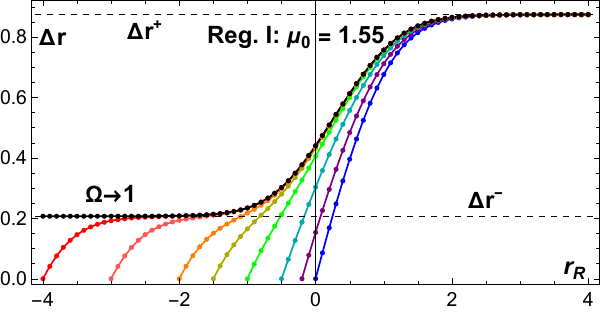}
\includegraphics[width=0.47\textwidth]{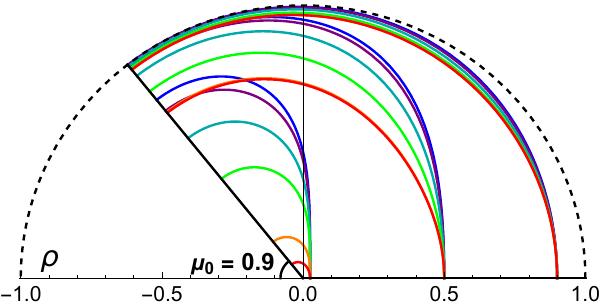}
\qquad
\includegraphics[width=0.47\textwidth]{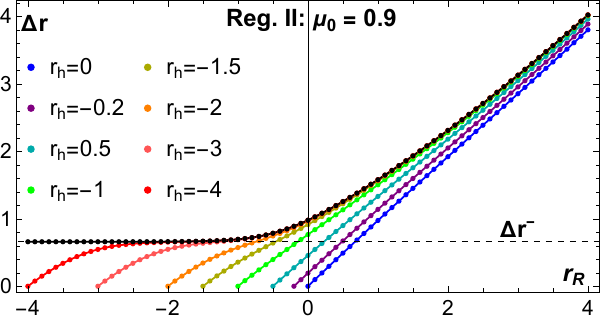}
\includegraphics[width=0.47\textwidth]{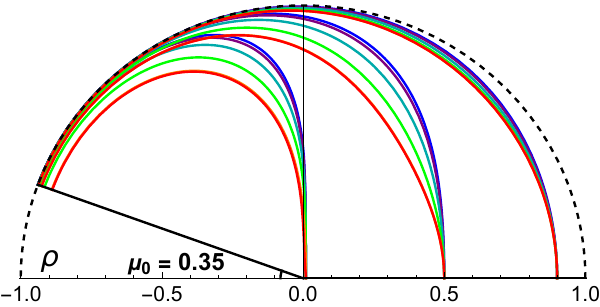}
\qquad
\includegraphics[width=0.47\textwidth]{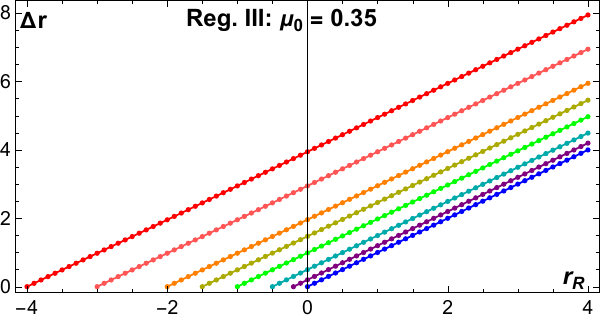}
\caption{Behavior of island surfaces in the three regimes \textbf{I}, $\mu_0>\mu_c$, at the top, \textbf{II}, for $\mu_e<\mu_0<\mu_c$, middle row, and \textbf{III}, $\mu_0<\mu_e$, bottom row. Curves with the same color are associated with the same value of $\Omega$, which increases ($r_h$ decreases) from the static solution (blue curve $\Omega=0,$ $r_h=0$), to the near extremal case (red curve $1-\Omega=5\cdot10^{-6},$ $r_h=-4$). Black lines correspond to the extremal solution. 
}
\label{fig:5DRotatingIslands}
\end{figure}

\begin{figure}[h!]
\centering
\includegraphics[width=0.49\textwidth]{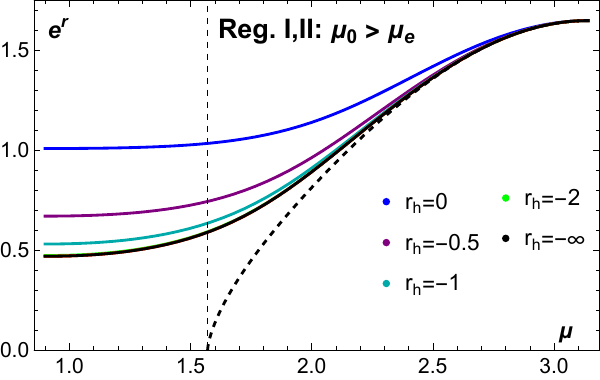}
\includegraphics[width=0.49\textwidth]{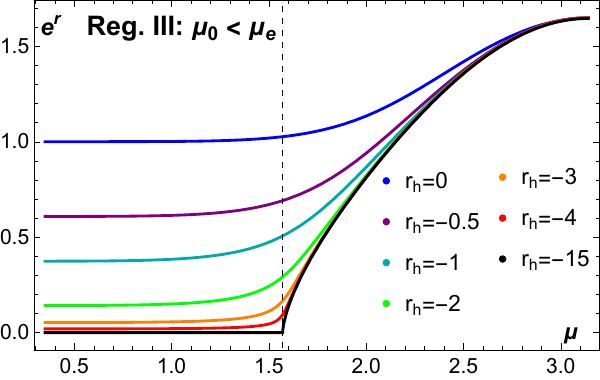}
\caption{Behavior of island surfaces in the extremal limit at fixed $r_R$ in regimes (\textbf{I,II}), here with $\mu_0 = 0.9>\mu_e$ (left), and \textbf{III} with $\mu_0=0.35<\mu_e$ (right). We show the radial coordinate $e^r$ so that the horizon is located at $e^r=0$ at extremality. The black dashed curve corresponds to the HM surface at $r_h=-15$, very close to extremality. In regime \textbf{III} this curve  essentially coincide with the corresponding island surface (solid black).}
\label{Fig:5DIslandSurfaceExtremalLimit}
\end{figure}

Examples of island surfaces are shown for various combinations of the $\mu_0,r_h$ and $r_R$ parameters in Figs. \ref{fig:5DRotatingIslands} (left), and \ref{Fig:5DIslandSurfaceExtremalLimit}.

We find that rotation has a ``repulsion" effect on island surfaces.
If we study families of surfaces with fixed $\delta r=r_R - r_h$, in fact, as we increase rotation, we see that the surfaces' radial profile is pushed away from the black hole horizon (and closer to the defect). This effect is shown in Fig. \ref{fig:5DRotatingIslands} (left), where islands are parametrized by the radial coordinate $\rho = e^{-(r-r_h)}$, with the horizon at $\rho=1$ and the defect at $\rho=0$. This allows for a convenient visualization of the islands relative to the horizon.

Compared to the static case, our analysis shows that, when rotation is added, a new regime appears with qualitatively different properties of the surfaces, which we will discuss in details below. There are thus three main regimes of parameters, which we denote (\textbf{I,II,III}), that are reached by changing the KR brane angle $\mu_0$. The first regime is controlled by the critical angle $\mu_c$, that also governs the behavior of island surfaces in the static case. Its numerical value is again \eqref{eq:MucValue}, it coincides with the angle controlling the existence of island surfaces for empty Anti de Sitter, showing this is not affected by rotation. This strengthens its interpretation as an ``universal" property of the braneworld model. Regime \textbf{III}  is new, and it is separated from \textbf{II} by a new critical brane-angle, intrinsically related to rotation. We denote it as $\mu_e$ and we numerically determine its value to be $\mu_e \sim 0.52872$, lower than $\mu_c$ in \eqref{eq:MucValue}. This critical angle is also the value that governs the existence of islands for the extremal solution, and does not depend on the cylindrical black hole parameters. This parallel with the Anti de Sitter case highlights the importance of the extremal configurations in the characterization of extremal surfaces. 

The critical angle $\mu_c$ defines the region of existence of the atoll, in the static case, as discussed in the previous subsection.  Also in the rotating case, the atoll appears only after lowering $\mu_0<\mu_c$, as can be seen in Fig. \ref{fig:5DRotatingIslands} (right), from the divergence of $\Delta r$ as $r_R \to+\infty$. Although the atoll is present both in regimes \textbf{II} and \textbf{III}, we find that the properties of the critical anchor $r_A$ are affected by rotation and change across the new critical value $\mu_e$: for any $\mu_0$ in \textbf{II}, $r_A$ as a function of rotation is bounded by a minimum value $r_{A_{min}}$, while in \textbf{III} it is unbounded.

In Fig. \ref{Fig:5DIslandSurfaceExtremalLimit} we show the behavior of islands as they approach extremality and how this changes across the critical angle $\mu_e$. We fix $r_R$ and change rotation. We see that island surfaces approach the extremal limit continuously in regimes \textbf{I} and \textbf{II}, but in regime \textbf{III} islands do not exist for the extremal solution as they show a singular behavior. What the plot shows is that islands in regime \textbf{III} tend to lay onto the HM surface as rotation approaches exremality.

We discuss in more details the three regimes, with reference to Figs. \ref{fig:5DRotatingIslands} and \ref{Fig:5DIslandSurfaceExtremalLimit}.

\paragraph{Regime I ($\mu_0>\mu_c$):} Top row in Fig. \ref{fig:5DRotatingIslands}. In this regime both the extremal solution and empty AdS support islands. In the extremal solution they are given by the $r_h\to -\infty$ limit, at fixed $r_R$, of the surfaces found at finite-temperature, and are shown in black in Figs.  \ref{fig:5DRotatingIslands} and \ref{Fig:5DIslandSurfaceExtremalLimit}. In this regime, islands remain far from the horizon as extremality is approached at any fixed $r_R$.

At large $r_R$, islands behave like those of empty AdS, with $\Delta r$ reaching the constant value $\Delta r^+$. As $r_R$ decreases, rotation starts having a greater effect on the surfaces. Islands remain qualitatively similar to those of the static solution but the value of $\Delta r$ decreases due to the repulsion effect, which makes $r_L$ increase. This is evident if one shifts curves in \ref{fig:5DRotatingIslands} so that they all start from the origin, thus reach the boundary at the same $r_R-r_h$.

A new behavior appears as we approach extremality. As $r_R$ decreases, $\Delta r$ remains fixed at a constant value $\Delta r^-$ for a while before dropping to zero at $r_R = r_h$, forming an intermediate plateau. The size of the plateau increases as we approach extremality, extending indefinitely in the limit.  The formation of the plateau for sufficiently negative values of $r_R$, is reminiscent of the behavior at large $r_R$ which is related to the symmetries of empty AdS. This can be seen by looking at the area functional \eqref{eq:5DAreaTimeDepAreaFunctional} in the extremal limit and for $r\ll-1$:
\begin{equation}
A_\gamma^{extr} \sim_{r_R\ll -1} V_{cyl}\,\ell_5^3\, \sqrt{\frac{8 m}{3}} \int d\mu \, \frac{e^{r/2}}{\sin^3 \mu}\sqrt{1+r'^{\,2}}\,,
\label{eq:5DExtremalSmallrAreaFunct}
\end{equation}
where $m$ have been defined in \eqref{eq:CylindricalBHExtremalLimit}. As $r(\mu)$ is an increasing function of $\mu$, surfaces that start at $r_R\ll-1$ remain at $r(\mu)\ll-1$ and can be described using \eqref{eq:5DExtremalSmallrAreaFunct}.

A shift in $r$ results in a rescaling of \eqref{eq:5DExtremalSmallrAreaFunct} so that the equation of motion is left invariant. This explains the origin of the plateau. For near-extremal solutions $r_R$ is bounded by the horizon and the plateau forms when $r_R$ is large and negative but not too close to the horizon, where the induced area functional takes a different form than \eqref{eq:5DExtremalSmallrAreaFunct}.

\begin{figure}[]
\centering
\includegraphics[width=0.6\textwidth]{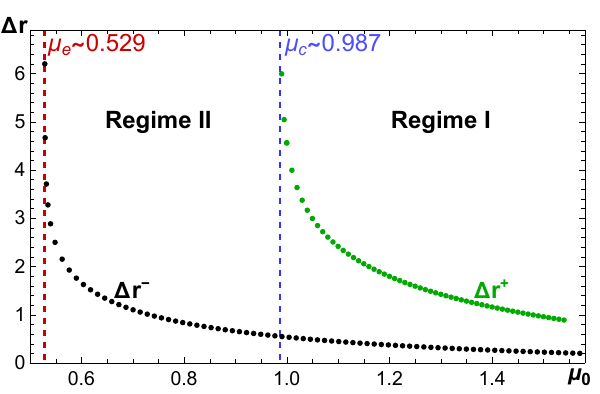}
\caption{Plots of the constant values $\Delta r^{\pm}$ as a function of $\mu_0$. $\Delta r^+$, ($\Delta r^-$) diverges as we approach $\mu_c$, ($\mu_e$) from above. Correspondingly, empty AdS and the extremal solution do not support islands below the associated critical angles. }
\label{Fig:5DDeltaPlusDeltaMin}
\end{figure}

\paragraph{Regime II ($\mu_e<\mu_0<\mu_c$):} Middle row in Fig. \ref{fig:5DRotatingIslands}.
Empty AdS does not support islands anymore, and the behaviour of island surfaces is modified at large $r_R$: $\Delta r^+$ diverges as we approach $\mu_c$ from above (see Fig. \ref{Fig:5DDeltaPlusDeltaMin}). Outside this region, the behavior of the surfaces remain the same as in Regime \textbf{I}. $\Delta r^-$ slowly increases as we decrease $\mu_0$ (see Fig. \ref{Fig:5DDeltaPlusDeltaMin}), meaning that the island surfaces of the extremal solution reach further down towards the singularity, but still remain at parametrically large distances from it, in the sense explained above.

\begin{figure}[]
\centering
\includegraphics[width=0.48\textwidth]{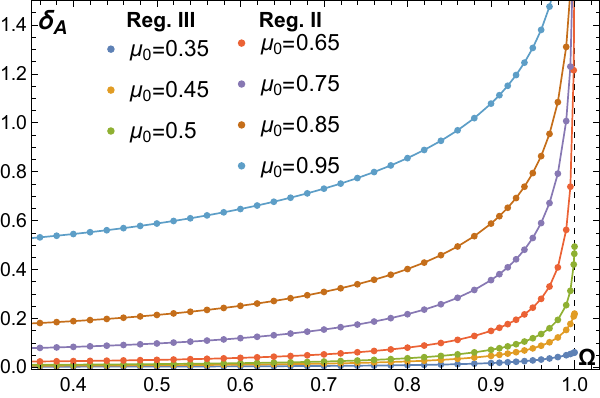}
\quad
\includegraphics[width=0.48\textwidth]{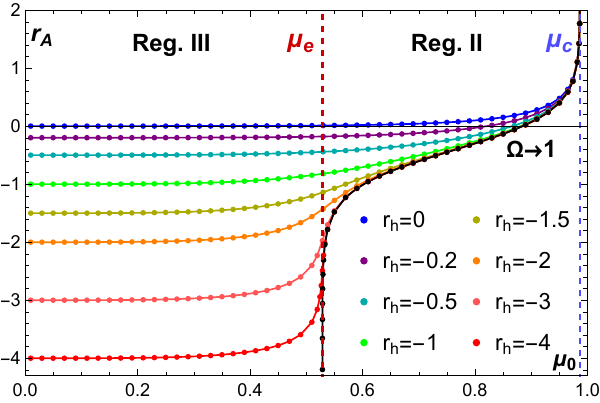}
\caption{Left: atoll size $\delta_A = r_A-r_h$ as a function of $\Omega$ for various brane angles. Right: critical anchor as a function of $\mu_0$, for different values of the rotation. The black datapoints correspond to the critical anchors of the extremal solution, which exists only in regime \textbf{II}.}
\label{Fig:5DAtollSizeCritAncor}
\end{figure}

We find that below $\mu_c$ the atoll forms, like in the static case, and $r_L \le r_A$. This induces the change of behavior of $\Delta r$ at large $r_R$ in Eq. \eqref{eq:IslandsDeltarBehavior}. Rotation influences the position of the critical anchor $r_A$ and consequently the quantity $\delta_A = r_A-r_h$, which is  invariant  under the rescaling \eqref{eq:PartialRescalingAreaFunctional}. The behaviors of $r_A$ and $\delta_A$ are shown in Fig. \ref{Fig:5DAtollSizeCritAncor}.

We find that increasing rotation at fixed $\mu_0$ makes the position of the critical anchor $r_A$ decrease slower than $r_h$, resulting in an increasing  $\delta_A$. Indeed, we find this to be a monotonically decreasing function as $\delta_A(r_h)$ (increasing function as $\delta_A(\Omega)$). As extremality is approached, $r_A(\Omega)$ tends to a finite value $r_A(1)=r_{A_{min}}$ (black points in Fig. \ref{Fig:5DAtollSizeCritAncor} right) which is also the smallest value for $r_A$ at the given value of $\mu_0$. As a consequence, $\delta_A$ diverges at extremality because a larger and larger portion of the brane is given access to the atoll by the receding horizon. 

This behavior of the critical anchor can also be seen using the $\rho$ coordinate (middle left plot in Fig. \ref{fig:5DRotatingIslands}). The anchor point $\rho_L\sim\rho_A$ for islands that start close to the defect, with $\rho_R\sim 0$, approaches the defect as rotation increases.

\paragraph{Regime III ($\mu_0<\mu_e$):} Bottom row in Fig. \ref{fig:5DRotatingIslands}. In this regime neither empty AdS, nor the extremal solution support islands. A diagnostic of this fact is  given by the divergence of $\Delta r^-$ as $\mu_e$ is approached from above, which we use to numerically determine the value of $\mu_e$. A more precise picture of why this is the case is obtained by looking at the extremal limit of the island surfaces at finite-temperature. This is shown in Fig. \ref{Fig:5DIslandSurfaceExtremalLimit} (right). 

As extremality is approached, $r_L$ follows the receding horizon decreasing without bound. This is in contrast to what happens in regimes (\textbf{I,II}) where $r_L$ remains bounded from below. At any small but non-zero value of the temperature, island surfaces display a very steep drop around $\mu=\frac{\pi}{2}$ in the $r(\mu)$ parametrization, before turning sharply to reach the brane at $r_L$. At extremality this point becomes a cusp, which always appears at $\mu=\frac{\pi}{2}$ for any value of the bath anchor point, which makes the would-be island surface singular. This can be seen in Fig. \ref{Fig:5DIslandSurfaceExtremalLimit} by looking at the black curve, which is very close to extremality. 

Additionally, upon including HM surfaces, we find that, as extremality is approached, they reach the horizon at $\mu=\frac{\pi}{2}$ coalescing with the corresponding island surfaces in this limit. This seems to indicate that, in regime \textbf{III}, the extremal solution only supports one class of extremal surfaces that reach the singularity at $\mu=\frac{\pi}{2}$.

This also has consequences on the properties of the critical anchor and the size of the atoll, which now approaches a fixed value at extremality, due to the fact that $r_A \sim r_h$ in that limit. 

Again, such behavior is also manifest using the $\rho$ coordinate. Surfaces that start close to the defect, with $\rho_R\sim0$, have a fixed value of $\rho_L \sim e^{-\delta_A(1)}$ at extremality, as can be seen in Fig. \ref{fig:5DRotatingIslands} (bottom left).

\subsubsection{Area difference}

\begin{figure}[]
\centering
\includegraphics[width=0.44\textwidth]{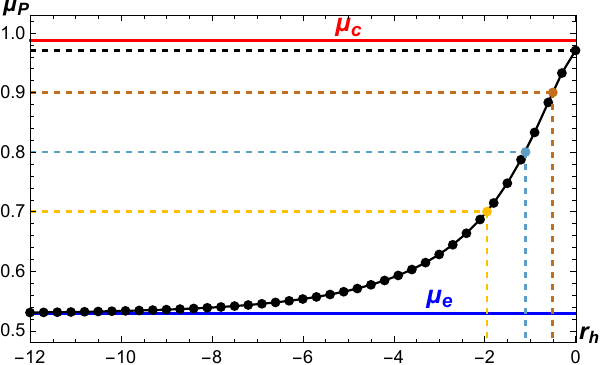}
\quad
\includegraphics[width=0.52\textwidth]{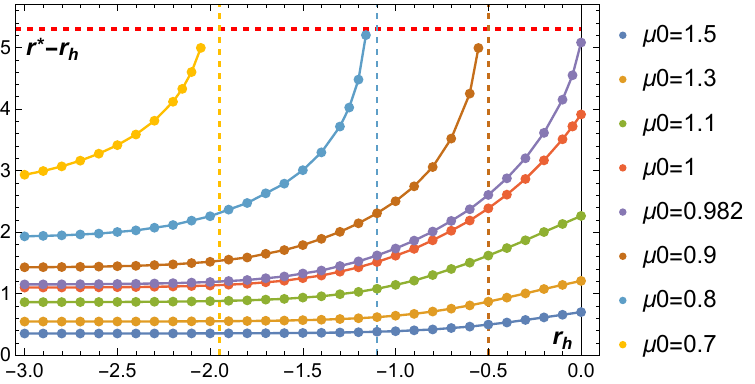}
\caption{Left: Page angle as a function of $r_h$, determined by studying extremal surfaces anchored at the defect. Right: $r^\star$ vs $r_h$ is shown for different brane-angles. As explained in the main text, the two estimations of $\mu_p$ seem to agree, despite us not being able to reliably reach the region with $r^\star-r_h > 5.3$ on the right plot.}
\label{Fig:5DrStarAndMupage}
\end{figure}

In order to study how rotation affects the entanglement phase structure of the dual BCFT, we need to take into account HM extremal surfaces. 
By studying the area difference \eqref{Area-diff}, we find that the size of the constant entropy belt $\mathcal{C}$ increases with rotation, as well as the range of values of $\mu_0$ for which it exists. Indeed, the Page angle becomes an increasing function of $r_h$, shown in Fig. \ref{Fig:5DrStarAndMupage} (left), with
\begin{equation}
    \mu_p(r_h \to -\infty) = \mu_e \,, \qquad \mu_p(0) \lesssim\mu_c\,.
\end{equation}
Interestingly, $\mu_p$ is bounded from below precisely by $\mu_e$, which therefore controls both the existence of islands in the extremal solution and the existence of the belt:  $\mathcal{C}$ can only exist above $\mu_e$, for a given rotation. It also follows that, as rotation increases, the range of parameters for which both the atoll and the belt $\mathcal{C}$ exists, given by $\mu_0 \in (\mu_p,\mu_c)$, widens.
 
\begin{figure}[]
\centering
\includegraphics[width=0.49\textwidth]{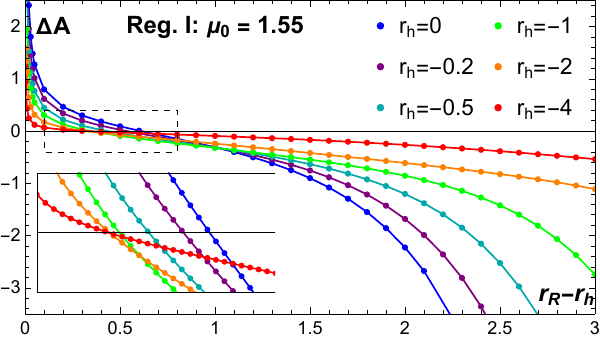}
\quad
\includegraphics[width=0.47\textwidth]{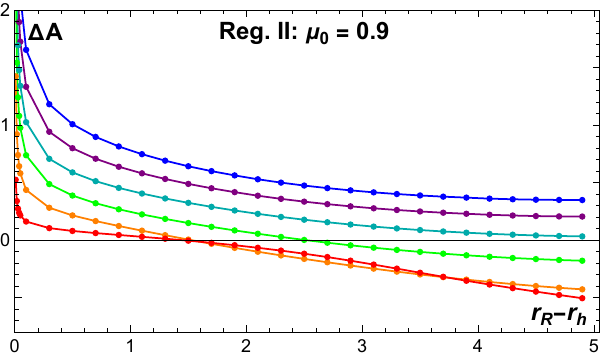}
\includegraphics[width=0.49\textwidth]{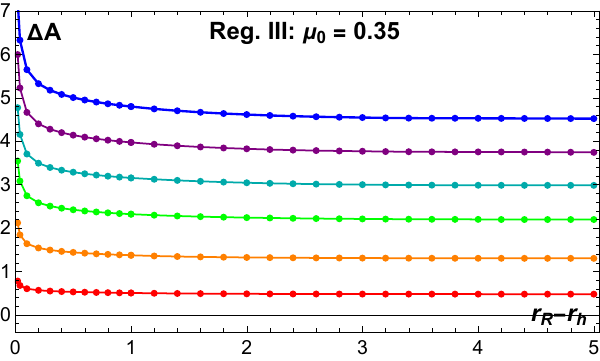}
\caption{Initial time area difference as a function of $r_R-r_h$ for different values of $\mu_0$ in the three regimes, and different values of the rotation (curves with different colors). 
}
\label{Fig:5DAreaDifference}
\end{figure}

In Fig. \ref{Fig:5DAreaDifference} we show how $\Delta A$ behaves in the three regimes as a function of $r_R-r_h$, for different values of rotation. Formally, the area difference also depends on the mass density $M$ by virtue of \eqref{eq:AreaScaling}. We continue to work with $M=1/4$ fixed, noting that we are interested in finding when $\Delta A =0$, which is not affected by \eqref{eq:AreaScaling}.

As in the static case, $\Delta A$ is a decreasing function of $r_R$ and $\mu_0$ even as we turn on rotation. It diverges to positive infinity as $r_R$ approaches the horizon, due to the HM surface shrinking to a point there, and to negative infinity in regime \textbf{I} as $r_R\to+\infty$ due to the shrinking of the island surface \cite{Geng:2021mic}. In regimes (\textbf{II,III}), the formation of the atoll implies that islands do not shrink anymore as they approach the defect and indeed the negative divergence disappears in these cases.

The dependence of $\Delta A$ on $r_h$ is more complicated. We find that if $\Delta A$ is positive for some value of $r_h$, then, increasing the rotation while keeping fixed $r_R-r_h$ and $\mu_0$, makes $\Delta A$ decrease. This implies that $r^\star$ and the page angle $\mu_p$ both decrease with rotation, corresponding to an increasing belt size (which recall it is defined by $r_R>r^\star$ in the bath). $r^\star$ and $\mu_p$ are shown in Fig. \ref{Fig:5DrStarAndMupage}. 
 
The numerical determination of $r^\star$ is limited by the precision at which we can find the area difference, with our methods we can only reliably reach $r_R-r_h\sim 5.3$, hence we do not have access to the region above the dashed horizontal red line in the plot in Fig. \ref{Fig:5DrStarAndMupage}. With this limitation, we cannot determine the page angle using its definition in terms of $r^\star$ in Eq. \eqref{eq:PageAngleDefinition}.
We can instead resort to the other equivalent  definition of $\mu_p$ as the brane-angle for which $\Delta A=0$ for surfaces ending at the defect, at a given value of $r_h$. To find such surfaces we use the shooting method from the defect, discussed in \cite{Karch:2023ekf}. This allows to determine $\mu_p$ by directly comparing the areas of the extremal surfaces that anchor exactly at the defect. 

The values of $\mu_p$ determined in this way seem to agree with the (rough) estimates that one can give by looking at the plot of $r^\star$ in Fig. \ref{Fig:5DrStarAndMupage} (right).  For $\mu_0 \le \mu_p(0)$, the $r^\star$ curves develop a vertical asymptote at a given value of $r_h=\hat r_h$. This defines $\mu_0$ as the page angle at $\hat r_h$. The dashed vertical lines show the position of the asymptote for $\mu_0 = \{0.9,0.8,0.7\}$ as estimated by the defect analysis (same-color points on the left plot), and seem to agree with the rough estimate of the position of the asymptote given by the $r^\star$ curves.

 Finally, note that the behavior of islands below $\mu_e$ can explain why a constant entropy belt cannot be found in this regime. Below $\mu_e$ the smallest values for the area difference are obtained for the near-extremal solutions, as we are well below the page angle for solutions far from extremality and $\Delta A$ decreases with rotation when it is positive. As we discussed above, the islands of the near-extremal solutions essentially coincide with the corresponding HM surfaces up to $\mu=\pi/2$ where they sharply turn to reach the brane at $\mu_0$. This means that the only non-zero contribution for $\Delta A$ comes from the island surface area from $\mu_0$ and $\mu=\frac{\pi}{2}$, leading to a positive area difference. 
 
\subsection{Critical angles in general dimensions}

The braneworld construction we just discussed can be directly generalized to arbitrary number of $d+1$ (bulk) dimensions \cite{Chen:2020uac, Geng:2020fxl}. In this section we give some comments about the general $d$ case, focusing on key aspects like the value and number of critical brane-angles.

 As a starting point, the most general solution that generalizes the cylindrical black hole \eqref{eq:RotatingCylindricalBH} to $d$ dimensions has been found in \cite{Awad:2002cz}. The rotation group is $SO(d-1)$ and allows for $n=[(d-1)/2]$ independent rotations, where $[x]$ is the integer part of $x$, parametrized by parameters $\Omega_i$ related to the angular velocity along each direction. The boundary geometry contains an $(S^1)^n$ factor, while the remaining spatial directions, parametrized by $z^k$, can be taken either compact or non-compact. The metric reads (setting $\ell_{AdS_d} = \ell_{\phi_i} = 1$):
\begin{align}
    ds^2 = \frac{dr^2}{b_d(r)}+ \frac{e^{2r}}{\Xi^2}\Biggl[&-b_d(r)\biggl(dt-\sum_{i=1}^n \Omega_i\,d\phi_i\biggr)^2+\sum_{i=1}^n\left(\Omega_i\,dt-d\phi_i\right)^2 \notag\\
    &-r^2\sum_{i<j}^n\left(\Omega_i\,d\phi_j-\Omega_j\,d\phi_i\right)^2\Biggr]+e^{2r}dz^kdz^k\,,
    \label{eq:GeneraldimBH}
\end{align}
where
\begin{equation}
    b_d(r) = 1-e^{-(d-1)(r-r_h)}\,, \qquad \Xi^2 = 1-\sum_i\Omega_i^2\,, \qquad \phi_i \sim\phi_i+2\pi\,,
\end{equation}
with the constraint $\Xi^2<1$ which implies $\sum_i\Omega_i^2 <1$. The extremal limit is achieved for $\Xi\to0$ and $r_h\to-\infty$ as before. The black hole solution \eqref{eq:GeneraldimBH}  has one event horizon at $r=r_h$ and it can be checked to be locally, but not globally, isometric to the d-dimensional static black brane solution, in complete analogy with the 4d cylindrical black hole. Its temperature is given by
\begin{equation}
    T = \frac{(d-1)\,\Xi\,e^{r_h}}{4\pi}\,.
\end{equation}

The embedding of the black hole \eqref{eq:GeneraldimBH} in the braneworld model is again achieved via the black string metric \eqref{eq:5DBlackStringGeometry}, which is now a d+1-dimensional metric, with $ds^2_{cyl}$ given by \eqref{eq:GeneraldimBH} and a KR brane embedded at $\mu=\mu_0$. The geometry is still captured by Fig. \ref{Fig:BlackStringGeometry}. 

Extremal surfaces are derived similarly as before. Concentrating on islands, we consider the embedding $r=r(\mu), t=t_0$ and look for solutions that minimize the induced area functional:
\begin{equation}
    A_\gamma =V_{d-2}\,\ell_{d+1}^{d-1} \int_{\mu_0}^\pi d\mu\,\frac{e^{(d-2)r}}{(\sin\mu)^{d-1}}\sqrt{\frac{1+(\Xi^2-1)b_d(r)}{\Xi^2}\left(1+\frac{r'^{\,2}}{b_d(r)}\right)}\,.
    \label{eq:GeneraldAreaFunctional}
\end{equation}
Note that \eqref{eq:GeneraldAreaFunctional} depends on the rotation parameters $\Omega_i$ only through $\Xi$. This implies that, as long as extremal surfaces are concerned, their qualitative properties only depend on the ``overall" rotation measured by $\Xi$ and not on its amount on each specific direction\footnote{We may have expected this result as the solutions \eqref{eq:GeneraldimBH} with multiple rotations are related to one another by local coordinate transformations that do not change the form of the induced area functional.}. Geometries with the same value of $\Xi$ but different amount and number of rotations have the same extremal surfaces. In particular, this implies that as we turn on more than one rotation, no new critical behaviors emerge and the physics of extremal surfaces is still governed by the same critical parameters of the single-spinning solution, which we find to be given by $\mu_0$ and $\mu_e$ in all dimensions $d>3$. This strengthens the interpretation of these critical values as intrinsic properties of the braneworld model that control its entanglement structure. 

\begin{figure}[]
\centering
\includegraphics[width=0.6\textwidth]{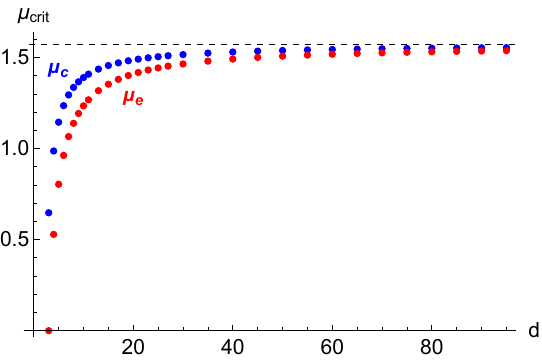}
\caption{Critical brane-angles in different dimensions.}
\label{fig:CriticalAnglesGeneralD}
\end{figure}

We show in Fig. \ref{fig:CriticalAnglesGeneralD} how the critical angles depend on the number of dimensions (the same plot for $\mu_c$ alone have been given in \cite{Geng:2020fxl}). Both $\mu_c$ and $\mu_e$ increase with $d$ and reach $\pi/2$ in the limit. For $d=3$ we find that $\mu_e=0$ implying that the extremal solution always supports islands. This can be understood by looking at the behavior of $\Delta r^-$, which in $d=3$ is found to be identically zero for all brane-angles while, for $d>3$, diverges at $\mu_0=\mu_e$. This happens because the ODE for islands in the $d=3$ extremal solution  takes a particularly simple form in the region with $r\ll-1$:
\begin{equation}
   r''- 2\,r'\cot\mu\,\bigl(1+r'^{\,2}\bigr)=0\,,
   \label{eq:3dExtremalNegativerODE}
\end{equation}
which is trivially solved by constant solutions $r(\mu)=r_R$. It follows that the islands of the $d=3$ extremal solution are characterized by a value of $\Delta r$ that tends to zero as $r_R$ decreases, implying $\Delta r^-\equiv0$. This holds for all values of $\mu_0$ and also implies that islands always exist in the $d=3$ extremal solution. This is not the case in $d>3$, where the ODE analogous to \eqref{eq:3dExtremalNegativerODE} still depends only on $r''$ and $r'$ but acquires an additional constant term, leading to more complicated solutions. Ultimately, this difference is what makes solutions of the ODE cease to exist below $\mu_e$ in $d>3$, which is detected by the divergence of $\Delta r^-$ at $\mu_0=\mu_e$.

\section{Embedding in Type IIB}
\label{Sez:10dModel}
In order to have microscopic control on the physical set-up, it is important to be able to connect it to a microscopic construction in a string theoretic framework. The AdS$_5$/AdS$_4$ geometry dual to 3d conformal interfaces in $\mathcal{N}=4$ Super Yang-Mills is known since the type IIB model of \cite{DHoker:2007zhm,DHoker:2007hhe,Aharony:2011yc,Assel:2011xz}, and has been first exploited for double holography in \cite{Uhlemann:2021nhu} and then elaborated in \cite{Karch:2022rvr}. In fact, it is possible to embed in 10d solutions of 4d Einstein gravity with negative cosmological constant \cite{Uhlemann:2021nhu}, possibly allowing the dilaton to vary \cite{Demulder:2022aij} or consider binary black holes \cite{Deddo:2023oxn}.

The focus of our analysis is the cylindrical black hole embedded in type IIB theory, and how the rotation parameter affects the entanglement surfaces. Because of the cylindrical topology the dual theory lives on $\mathbb{R}^{1,2}\times S^1$. Our study shows a similar structure to the 5d setting of the previous section, with the appearance of a second critical behavior related to the extremal limit of the cylindrical black hole. 
The qualitative agreement between the 5d-10d double holographic realizations could be expected, although the effect of the internal geometry in a generic extremization procedure makes a quantitative match out of reach. This is in line with the non-rotating black hole/black brane analysis of \cite{Uhlemann:2021nhu}, and with recent studies of relations between top-down and the bottom-up models \cite{Harvey:2025ttz}.

\subsection{10 Dimensional geometry}
We begin by quickly reviewing the 10d supergravity solutions modeling black holes coupled to asymptotic baths, and providing the microscopic realization of the 5d backgrounds considered above. 
We are interested in a single AdS$_5$ asymptotic region, which is consistent with the description of a black hole coupled to a non-gravitating bath, so the space is capped-off on one side of the AdS$_4$ interface. 

The geometry is a fibration of $AdS_4\times S^2 \times S^2$ over a base space which is a Riemann surface $\Sigma$ with disk topology \cite{DHoker:2007zhm,DHoker:2007hhe,Assel:2011xz}
\begin{equation}
    ds^2 = f_4^2\,ds^2_{[AdS_4]}+f_1^2\,ds^2_{[S_1^2]}+f_2^2ds^2_{[S_2^2]}+4\rho^2\,|dz|^2\,,
    \label{eq:10DMetric}
\end{equation}
where $ds_4^2$ and $ds^2[S_i^2]$ are the metrics of unit AdS$_4$ and of the unit sphere. The complex coordinate $z$ parametrizes the Riemann surface $\Sigma$, which we choose to be the infinite strip
\begin{equation}
    \Sigma = \{ z\in \mathbb{C}| \,\text{Im}(z) \in [0,\pi/2]\}\,.
\end{equation}
The metric functions $\{f_4,f_1,f_2,\rho \}$ are fixed in terms of two harmonic functions $h_1,h_2$ on $\Sigma$, which contain all information on the microscopic geometry
\begin{align}
		f_4^8&=16 \frac{N_1 N_2}{W^2}\ ,
		\qquad 
		\rho^8 = \frac{N_1N_2 W^2}{h_1^4 h_2^4} \ ,
		\qquad
		f_1^8=16h_1^8\frac{N_2 W^2}{N_1^3} \ ,
		\qquad
		f_2^8 =16h_2^8 \frac{N_1 W^2}{N_2^3} \ ,
\end{align}
with
\begin{align} 
    W&= \partial h_1\bar\partial h_2 + \bar\partial h_1\partial h_2 = \partial\bar\partial (h_1 h_2) \ ,
    \\
    N_1&=2h_1 h_2|\partial h_1|^2-h_1^2 W \ ,
    \\
    N_2&=2h_1 h_2|\partial h_2|^2-h_2^2  W \ ,
\end{align}
and the dilaton is
\begin{align}
    e^{4\phi}=\frac{N_2}{N_1} \ ,
\end{align}
and will be a constant with the choice of our background below.
The 3-forms $H_3,F_3$ live in the internal geometry and only the 5-form $F_5$ has a term proportional to the AdS$_4$ worldvolume element and a one-form on $\Sigma$, supporting the AdS cosmological constant. Their explicit expressions can be found in \cite{DHoker:2007zhm,DHoker:2007hhe,Aharony:2011yc,Assel:2011xz}.

\begin{figure}[]
\centering
\includegraphics[width=0.48\textwidth]{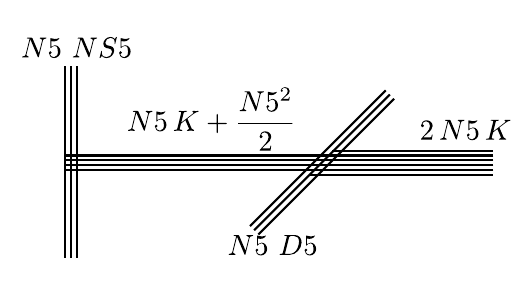}
\quad
\includegraphics[width=0.48\textwidth]{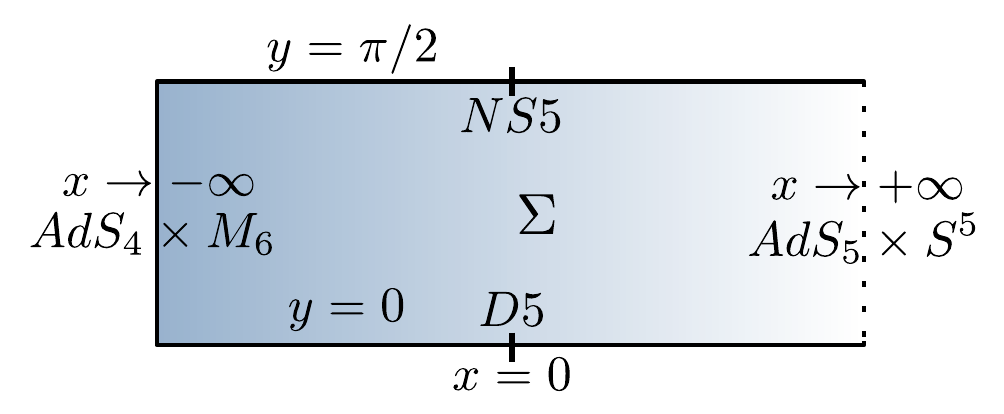}
\caption{Brane configuration and geometry associated to the choice of harmonic functions in \eqref{eq:10DHarmonicFunctions}.}
\label{fig:BraneConfig}
\end{figure}

For a background with a single $AdS_5\times S^5$ asymptotic region and branes localized at $Re(z)=0$ the harmonic functions read 
\begin{align}
    h_1 &= \frac{\pi\,\alpha'K}{4} e^z - \frac{\alpha' N_5}{4} \log \tanh \left(\frac{z}{2}\right) + c.c. \notag \ , \\
    h_2 &= -i\frac{\pi\,\alpha'K}{4} e^z - \frac{\alpha' N_5}{4} \log \tanh \left(\frac{i\pi}{4}-\frac{z}{2}\right) + c.c. \ .
    \label{eq:10DHarmonicFunctions}
\end{align}
The microscopic brane configuration associated to the functions \eqref{eq:10DHarmonicFunctions} is shown in Fig. \ref{fig:BraneConfig} and consists of $2N_5K$ semi-infinite  $D3$-branes, half of which end on stacks of $N_5$ $D5$-branes and the other half on stacks of $N_5$ $NS5$-branes. Suspended between the five-branes there are additional $N_5^2/2$ $D3$-branes. The five-brane sources are visible in the metric \eqref{eq:10DMetric} and are associated with the singularities of $h_1,h_2$ at $z=\{0,0\}$ and $z = \{0,i\pi/2\}$. These 10d geometries  are dual to $\mathcal{N}=4$ SYM on half space coupled to 3d $T_\rho^\sigma\left[SU(N)\right]$ theories on the boundary \cite{Aharony:2011yc,Assel:2011xz}. 

The ratio between 3d and 4d degrees of freedom in the dual BCFT is controlled by the parameter $\alpha = N_5/K$, that plays the same role as the brane-angle $\mu_0$ and also controls the graviton mass in the effective 4d theory \cite{Bachas:2018zmb,Demulder:2022aij}. It is possible to relate $\alpha$ and $\mu_0$ \textit{qualitatively} by
\begin{equation}
    \alpha \sim \mu_0^{-1}\, .
\end{equation}
Given the choice of harmonic functions in \eqref{eq:10DHarmonicFunctions}, with $z=x+i y$, $x\,,y\in \mathbb{R}$, one recovers an $AdS_5\times S^5$ region from \eqref{eq:10DMetric} in the asymptotic limit $x\to+\infty$, where $y$ acts as one of the angular coordinates of $S^5$ and $x$ becomes the radial coordinate of AdS$_5$, with AdS$_4$ slicing;\footnote{Indeed at large $x$ the 10d metric reduces to 
\begin{align}
    ds^2_{10D} = L_5^2\left[\cosh^2(c-x)ds^2_{AdS_4}+dx^2\right]+L_5^2\,ds^2_{S^5} \ ,
\end{align} with the part in square brackets being the metric of AdS$_5$ in AdS$_4$ slices and $c= \log (2\alpha/\pi)$. The $x$-coordinate is related to the $\mu$-coordinate in \eqref{eq:5DBlackStringGeometry} as $\cosh^{-1}(c-x) = \sin \mu$.} both AdS$_5$ and $S^5$ have radii $L_5^4 = 8\pi\alpha'^2N_5K$. The left asymptotic region reached at $x\to-\infty$ is a regular point of the geometry that, here, is approximated by $AdS_4\times M_6$. 
From the 10d point of view, the full geometry can be visualized as a  4d $AdS_4\times M_6$ ``bag" region, supporting the five-brane charges, with a noncompact ``opening'' along the $x$ coordinate  \cite{Bachas:2018zmb, Demulder:2022aij} producing a throat that connects it to a five-dimensional $AdS_5\times S^5$ region. If we think of the  $AdS_5\times S^5$ region representing the ``bath'', this geometry gives a string theory realization of the Karch-Randall braneworld, despite in this case a sharp distinction between a 4d and a 5d region cannot be made as in the localized, thin-brane picture \cite{Karch:2022rvr}. Indeed, the top-down construction yields a ``fat-brane" AdS$_4$ region \cite{Bachas:2018zmb}, which offers nonetheless a convenient microscopic realization of a doubly-holographic model.  Accordingly, we will find that the results we obtained in the bottom-up model for the cylindrical black hole remain qualitatively valid also in the top-down construction. 

Finally, the embedding of the cylindrical black hole is given by replacing the
AdS$_4$ factor in \eqref{eq:10DMetric} with the cylindrically rotating black hole \eqref{eq:RotatingCylindricalBH} with $\ell_\phi = \ell_4=1$ \cite{Uhlemann:2021nhu}.
The resulting metric still satisfies the Type IIB supergravity equations of motion as \eqref{eq:RotatingCylindricalBH} is a solution of 4d pure Einstein gravity with negative cosmological constant. This is the 10d geometry which we are going to use to derive the extremal surfaces.

\subsection{10d Extremal surfaces}
 As before, the extremal surfaces we are interested in should be derived using the HRT prescription, hence considering general codimension-2 surfaces associated with a subregion $\mathcal{R}$ in the dual BCFT defined at $t = const.$ and $r\le r_R$. In practice we find also in this case that, once we impose appropriate boundary conditions, the simpler RT prescription is consistent with the more general HRT. A key difference from the 5d case is that now the two prescriptions are not equivalent, using the RT prescription gives consistent solutions that, however, may not be the unique ones satisfying the given boundary conditions. We will not try to find if more general solutions exist and just consider the time-independent ones.

\subsubsection{Island surfaces}
We consider codimension-2 surfaces that wrap both $S^2$'s and the cylinder coordinates $\phi,z$, described by the  embedding functions:
\begin{equation}
t=t(x,y)\,,\qquad r=r(x,y)\,,
\end{equation}
where remember that $z=x+iy$.

The induced metric reads:
\begin{align}
    ds^2_{\gamma} = f_4^2\,e^{2r}\,dz^2+f_1^2\,ds^2_{S_1^2}+f_2^2\,ds^2_{S_2^2}&+f_4^2\Bigl[F_x\,dx^2+F_y\,dy^2+F_\phi\,d\phi^2  \notag \\ 
    &+2G_{xy}\,dxdy+2G_{x\phi}\,dxd\phi+2G_{y\phi}\,dyd\phi\Bigr]\,,
\end{align}
where:
\begin{align}
     F_i &= \frac{4\rho^2}{f_4^2}-e^{2r}\frac{b(r)-\Omega^2}{1-\Omega^2}(\partial_it)^2+\frac{(\partial_ir)^2}{b(r)}\,, \qquad\quad F_\phi =  e^{2r}\frac{1-\Omega^2 b(r)}{1-\Omega^2}\,, \notag \\
    \quad G_{i\phi} &= e^{2r}\,\frac{\Omega\left(b(r)-1\right)}{1-\Omega^2} \,\partial_it\,, \qquad\quad
    G_{xy} =-e^{2r}\frac{b(r)-\Omega^2}{1-\Omega^2} (\partial_xt)\, (\partial_yt)+\frac{(\partial_x r)\,(\partial_yr)}{b(r)}\,.
    \label{eq:10DInducedMetricFUnctions}
\end{align}
The area functional reads $A_\gamma = V_{cyl}\,V_{S_1^2}\,V_{S_2^2}\,S_\gamma\,,$
with:
\begin{equation}
    S_\gamma = \,\int dxdy\,e^r\,f_4^4f_1^2f_2^2\sqrt{F_xF_yF_\phi-F_\phi G_{xy}^2-F_x G_{y\phi}^2-F_y G_{x\phi}^2+2G_{xy}G_{x\phi}G_{y\phi}}\,.
\end{equation}
By virtue of the stationarity of the spacetime, \eqref{eq:10DInducedMetricFUnctions} is a functional that only depends on the derivatives of $t(x,y)$. This implies that the equation for $t$ takes the form of a conservation equation like in the 5d case, calling $S_\gamma = \int L_\gamma$:
\begin{equation}
    \delta_t S_\gamma = \partial_x\left[ \frac{\partial L_\gamma}{\partial\,(\partial_xt)}\right]+\partial_y\left[ \frac{\partial L_\gamma}{\partial\,(\partial_yt)}\right] =0\,,
    \label{eq:10DTimeEoM}
\end{equation}
one can check that a trivial solution of \eqref{eq:10DTimeEoM} that satisfies the boundary conditions (to be discussed below) is given by $t = const.$ and hence by the usual constant-time hypersurfaces one obtains with the RT prescription.

This is analogous to what happens in 5d, but the argument leading to $t=const.$ is weaker as Eq. \eqref{eq:10DTimeEoM} is a conservation equation in \textit{two} variables rather than one and cannot be integrated like Eq. \eqref{eq:5DTimeEOM}, leading to a first order equation for $t$. Ultimately this leads to a weaker constraint on $t$ once we impose the boundary conditions. Indeed, remember that just requiring $t'(\mu_0)=0$ in 5d automatically implies $t'(\mu) \equiv0$ by virtue of \eqref{eq:5DTimeEOM}, while now one should solve the full second order PDE \eqref{eq:10DTimeEoM}. 

With this in mind, we will ignore more general time-dependent solutions, and just look for fixed-time ones,  parametrized by just one function $r=r(x,y)$ as in \cite{Uhlemann:2021nhu}. Then, the area functional simplifies to:
\begin{equation}
    S_{\gamma} = 32\,L_5^8\int\,dxdy\,f\,e^{2r}\sqrt{\frac{1-\Omega^2b(r)}{1-\Omega^2}\bigl(1+g\,(\nabla r)^2\bigr)}\,,
    \label{eq:10DIslandAreaFunctional}
\end{equation}
where:
\begin{equation}
    L_5^8\,f= \frac{1}{8}\,f_4^2f_1^2f_2^2\rho^2 = |h_1h_2W|\,,\qquad g = \frac{1}{4}\frac{f_4^2}{\rho^2\,b(r)}= \frac{1}{2b(r)}\left|\frac
    {h_1h_2}{W}\right|\,.
    \label{eq:10DDefinitionOfFandG}
\end{equation}
With these definitions, the only dependence of $f,g$ on  microscopic parameters is through $\alpha=N_5/K$. The area functional $S_\gamma$ also depends on the microscopic parameters through $L_5$, which is a multiplicative factor, which we can express as $L_5^4 = 8\pi\alpha'^2 \alpha K^2$.  $K$ appears as an overall parameter which does not affect the PDE, we will fix $K=1$ to construct numerical surfaces in the next section.
Furthermore, the area functional \eqref{eq:10DIslandAreaFunctional} is simply rescaled under the transformation \eqref{eq:PartialRescalingAreaFunctional} that shifts $r_h$, analogously to the 5d case. Again, this implies that extremal surfaces display new qualitative features only as we vary $\Omega$.

The PDE for $r(x,y)$, obtained from the variation of $S_{\gamma}$, is only slightly modified from the non-rotating case:
\begin{equation}
    \frac{\delta L_{\gamma}}{L_{\gamma}} = \frac{1}{1+g(\nabla r)^2}\left[2-\nabla(g\nabla r)+\frac{1}{2}g\nabla r\cdot\nabla \ln \left({\frac{1+g(\nabla r)^2}{b(r)f^2}}\right) -\frac{3\Omega^2}{2}\frac{1-b(r)}{1-\Omega^2 b(r)}{}\right]=0\,.
    \label{eq:10DIslandPDE}
\end{equation}

\subsubsection*{Island boundary conditions}
We need to impose boundary conditions to ensure regularity of the surface $\gamma$ on the boundaries of the strip $\Sigma$. Near $y=0$, the sphere 
$S_2^2$ collapses with $f_2^2\sim4y^2\rho^2$, while the other functions reach a nonzero value. The induced metric on $\gamma$ near $y=0$ takes the form:
\begin{align}
    ds^2_\gamma \sim4\rho^2&\left(dx^2+dy^2+y^2\,ds^2_{S_2^2}\right)+f_4^2\Biggl[\frac{(\partial_xr\,dx+\partial_yr\,dy)^2}{b(r)}-\frac{b(r)-\Omega^2}{1-\Omega^2}\bigl(\partial_xt\,dx+\partial_yt\,dy\bigr)^2
    \notag\\
    & +2\frac{\Omega\bigl(b(r)-1\bigr)}{1-\Omega^2}\bigr(\partial_xt\,dx+\partial_yt\,dy\bigl)\,d\phi\Biggr]+\cdots
 \end{align}
where the dots indicate other regular terms as $y\to0$.
Similarly to \cite{Uhlemann:2021nhu}, in order to avoid a conical singularity at $y=0$, the terms proportional to $dy^2$ inside the square bracket must vanish. This condition is trivially satisfied by requiring that
\begin{equation}
    \partial_yr|_{y=0} = \partial_yt|_{y=0} =0\,.
\end{equation}
A similar reasoning applies for $y=\frac{\pi}{2}$.

In imposing the boundary conditions at $y=0,\pi/2$, one should recall the presence of the five-brane sources at $x=0$, which are associated with singularities in the metric functions. As shown in \cite{Uhlemann:2021nhu} these singularities are also present in the extremal surfaces we consider here which, at zero temperature, reach the Poincaré horizon at $r\to-\infty$ at the location of the five-brane sources. At finite temperature, this behavior is regulated by the horizon. As discussed in \cite{Uhlemann:2021nhu} these singularities are ``mild", in the sense that the area of the extremal surfaces remain finite as the singularities are approached. For this reasons, at the location of the five-brane sources we do not impose any regularity condition.

As $x\to-\infty$ the space closes off smoothly, approaching a regular point of the geometry. This implies that $t(x,y)$ and $r(x,y)$ should be independent of $y$ in this limit to have a regular surface. Moreover, in the $x\to-\infty$ limit, the metric functions behave like: (introducing $v = 2e^x$, and taking $v\to0$):
\begin{equation}
    \qquad f_1^2\sim L^2v^2\sin^2y\,,\qquad f_2^2\sim L^2v^2\,\cos^2y\,,\qquad 4\rho^2\sim L^2v^2\,,
\end{equation}
where $L$ is fixed in terms of $N_5$ and $K$ and $f_4$ reaches a finite limit.
The induced metric on $\gamma$ then reads:
\begin{align}
    ds^2_\gamma \sim &L^2\left(dv^2+v^2dy^2+v^2\sin^2y^2\,ds^2_{S_1^2}+v^2\cos^2y\,ds^2_{S_2^2}\right)+f_4^2\Biggl[\frac{(\partial_vr\,dv+\partial_yr\,dy)^2}{b(r)}\notag\\
    &-\frac{b(r)-\Omega^2}{1-\Omega^2}\bigl(\partial_vt\,dv+\partial_yt\,dy\bigr)^2
     +2\frac{\Omega\bigl(b(r)-1\bigr)}{1-\Omega^2}\bigr(\partial_vt\,dv+\partial_yt\,dy\bigl)\,d\phi\Biggr]+\cdots
\end{align}
again, a conical singularity at $v=0$ is avoided if the quantity in the square brackets proportional to $dv^2$ vanishes, this condition is trivially satisfied by requiring that: 
\begin{equation}
    \lim_{x\to-\infty} e^{-x}\partial_xr = \lim_{x\to-\infty}e^{-x}\partial_x t = 0\,.
\end{equation}
Finally, in the bath region at $x\to+\infty$ we impose Dirichlet boundary conditions in the form:
\begin{equation}
    r(x,y) \xrightarrow[x\to+\infty]{}r_R\,,\qquad t(x,y)\xrightarrow[x\to+\infty]{} t_R=0\,.
\end{equation}
Notice that the hypersurfaces with $t(x,y)\equiv t_R$ trivially satisfies the above boundary conditions.

\subsubsection{HM Surfaces}
To discuss Hartman-Maldacena surfaces we again use the tortoise radial coordinate $u$ as in \eqref{eq:RotatingCylBHTortoise}. We parametrize the surfaces with $x=x(u,y)$ and set $t=0$ from the start. 

The induced metric reads:
\begin{align}
    ds^2 = &e^{2r_h} \cosh^{4/3}\left(\frac{3u}{2}\right)f_4^2\left[\frac{1-\Omega^2b(u)}{1-\Omega^2}d\phi^2+dz^2\right]+f_1^2\,ds_{S_1^2}^2+f_2^2ds_{S_2^2}^2 \notag \\
    &+ \left[f_4^2+4\rho^2(\partial_ux)^2\right]\,du^2+ 4\rho^2\left[ \bigl(1+(\partial_y x)^2\bigr)dy^2+2(\partial_u x)(\partial_y x)\,du\,dy\right]\,,
\end{align}
with associated area functional given by
\begin{equation}
    S_\gamma =  32\,L_5^8 \int\,dudy\,e^{2r_h}\,f\,\cosh^{4/3}\left(\frac{3u}{2}\right)\sqrt{\frac{1-\Omega^2b(u)}{1-\Omega^2} \biggl[\tilde g\,\bigl(1+(\partial_yx)^2\bigr)+(\partial_ux)^2\biggr]}\,,
    \label{eq:10DInducedAreaHM}
\end{equation}
where $\tilde g = b(u)\,g$ and $f,g$ have been defined in \eqref{eq:10DDefinitionOfFandG}.
The regularity of the surface on the boundaries $y=\{0,\pi/2\}$ of $\Sigma$ requires us to impose
\begin{equation}
    \partial_y x(u,0) = \partial_y(u,\pi/2) = 0\,,
\end{equation}
the Dirichlet condition on the asymptotic bath region becomes:
\begin{equation}
    \lim_{u\to u(r_R)} x(u,y) = +\infty\,,
\end{equation}
while at the horizon at $u=0$ we impose the Neumann boundary condition
\begin{equation}
    \partial_u x(0,y) = 0\,.
\end{equation}
With these boundary conditions, the HM surfaces cross the horizon at $t=0$ along a curve $x_h(y) = x(0,y)$ that is determined numerically by solving the PDE.

\subsection{Numerical solutions}
We now proceed to study the effects of rotation on the extremal surfaces, which we find by numerically solving the associated PDEs for $r(x,y)$ or $x(u,y)$. We again fix $M=1/4$ and scan for different values of the bath anchor point $r_R$ for different combinations of $\alpha$ and rotation parameter (again we are mainly going to use $r_h$).

To do so we use the relaxation method employed in \cite{Uhlemann:2021nhu,Demulder:2022aij}. We introduce a fictitious time-dependence on the embedding functions: $r(x,y,\tau)$, $x(u,y,\tau)$ and let the surfaces dynamically evolve according to (considering e.g. island surfaces)
\begin{equation}
    \delta_\tau r(x,y,\tau) = -L_\gamma^{-1}\frac{\delta L_\gamma}{\delta r(x,y,\tau)}\,,
    \label{eq:10DRelaxationEquation}
\end{equation}
where the rhs in the above equation is given by Eq. \eqref{eq:10DIslandPDE} with $r(x,y)\to r(x,y,\tau)$. Starting from a trial surface $r(x,y,0)$ that satisfies the boundary conditions and letting $r(x,y,\tau)$ evolve according to \eqref{eq:10DRelaxationEquation} for a sufficiently large time $\tau_{max}\gg1$, makes the surface relax on a minimal area equilibrium configuration $r(x,y,\tau_{max})$ which gives an approximated solution to \eqref{eq:10DIslandPDE}. The quality of the approximation can be estimated by the residual function
\begin{equation}
    R = \left|L_\gamma^{-1}\frac{\delta L_\gamma}{\delta r(x,y,\tau)}\right|_{r = r(x,y,\tau_{max})}\,.
\end{equation}

To solve Eq. \eqref{eq:10DRelaxationEquation} we use the finite difference method at second order. The domain $\Sigma$ is first compactified by the change of coordinate $\xi = \tanh x \in (-1,1)$ and then discretized into a regular lattice. Then, the partial differential equation \eqref{eq:10DRelaxationEquation} is turned into a set of coupled first order differential equations that are solved numerically with Mathematica \cite{Mathematica}. We generally use a lattice with a total of $O(1000)$ points and $\tau_{max}=1000$, which we find to give qualitatively good results for values of $\alpha$ and $r_R$ that are not too large.
Higher resolutions are needed as $\alpha$ and $r_R$ increase, requiring lattices with up to $O(100)$ points along each directions.

As before, in the limit $r_h \to 0$ we recover the results of \cite{Uhlemann:2021nhu}, which agree qualitatively with the behavior of the 5d model, discussed in the previous section. Also when rotation is added, we still find qualitative agreement between the properties of island surfaces in the top-down and the bottom-up constructions. However, due to the increasing numerical complexity of the differential equation, it is not possible to have numerical control over a large range of parameters in the 10d setup. In particular, it becomes very difficult to construct surfaces anchored at large values of $r_R$.  

\begin{figure}[]
\centering
\includegraphics[width=0.5\textwidth]{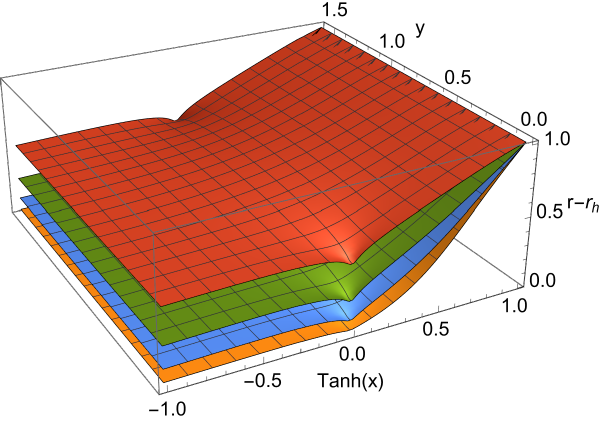}
\quad
\includegraphics[width=0.46\textwidth]{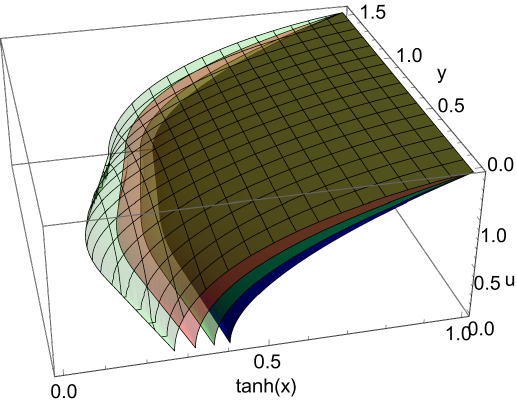}
\caption{Island surfaces (left) with $r_h=\{0,-0.5,-1,-4\}$, and HM surfaces (right) surfaces with $r_h = \{0,-0.5,-1,-1.82\}$, with increasing rotation from bottom to top. The other parameters are fixed to $\alpha=2$, $K=1$, $r_R=r_h+1$. The repulsion effect for island is evident from the plot, while HM surfaces are pushed towards the five-brane singularities.}
\label{fig:10DIslands}
\end{figure}

\subsubsection{Numerical study of extremal surfaces}

Examples of island and HM surfaces, for different rotation values, are shown in Fig. \ref{fig:10DIslands}.

\begin{figure}[]
\centering
\includegraphics[width=0.48\textwidth]{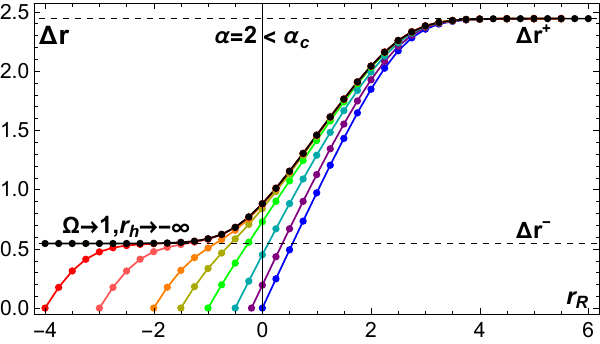}
\quad
\includegraphics[width=0.47\textwidth]{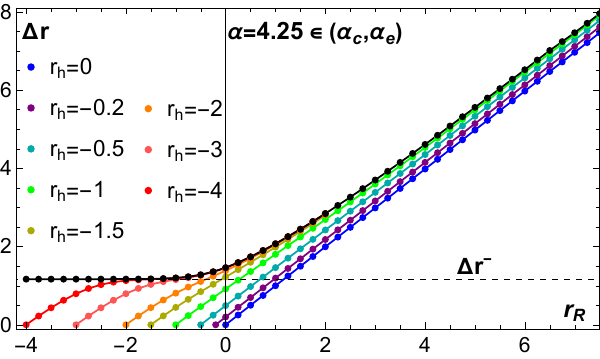}
\includegraphics[width=0.49\textwidth]{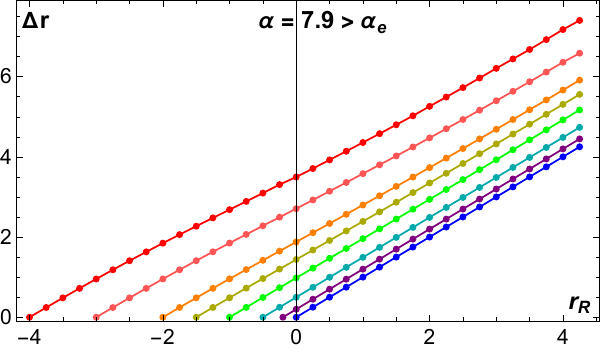}
\caption{$\Delta r$ vs $r_R$ in 10d in the three regimes bounded by the critical values: $\alpha_c \sim 4.0$ \cite{Uhlemann:2021nhu} and $\alpha_e \sim 7.8$. Comparing with the results in the 5d model in Fig. \ref{fig:5DRotatingIslands}, we find the same qualitative behavior for the island surfaces. For $\alpha>4$ it becomes difficult to find surfaces at large values of $r_R$, due to the limitations of our numerical method.}
\label{fig:10DDeltarvsrR}
\end{figure}

Again, $\Delta r$ is a useful quantity to study, in order to extract information about island surfaces. Plots showing $\Delta r$ vs $r_R$ are presented in Figure \ref{fig:10DDeltarvsrR}; by comparison with the 5d model in \ref{fig:5DRotatingIslands}, we find the existence of three different regimes of island surfaces also in the top-down model.

More precisely, there are two critical values of $\alpha$, that bound three regimes. They are found to be $\alpha_c\sim 4.0$, in agreement with \cite{Uhlemann:2021nhu}, and a new value, characteristic of the extremal rotating metric, $\alpha_e\sim 7.9$. We have mentioned previously that $\alpha$ can be related to the brane angle $\mu_0$ in the bottom-up model via $\alpha \sim \mu_0^{-1}$, so that $\mu_e<\mu_c$ while $\alpha_e>\alpha_c$. 
However, as we anticipated, the critical value $\alpha_e$ cannot be related quantitatively to $\mu_e^{-1}$, as the extra dimensions wrapped by the extremal surfaces actually modify the setup of the 5d braneworld, affecting the numerical critical value. It is not known, within this framework, how to quantitatively predict the value of $\alpha_e$ from $\mu_e$, knowing the uplift to 10 dimensions.

The behavior of islands in the three regimes follows the analogous discussion of the previous section. 

\begin{figure}[]
\centering
\includegraphics[width=0.6\textwidth]{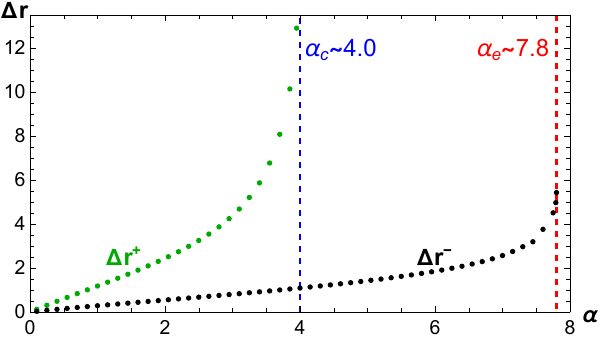}
\caption{Constant values $\Delta r^{\pm}$ as a function of $\alpha$. $\Delta r^\pm$ diverge as we approach the corresponding critical value $\alpha_{c/e}$.  }
\label{Fig:10DDeltaPlusDeltaMin}
\end{figure}

Below $\alpha_c$ (Regime \textbf{I}) empty AdS supports islands with a constant value of $\Delta r =\Delta r^+$ shown in Fig. \ref{Fig:10DDeltaPlusDeltaMin}. At finite temperature and rotation, islands tend to those of empty AdS at large $r_R$ and $\Delta r \to\Delta r^+$. In this regime the extremal solution supports islands, corresponding to the $r_h\to-\infty$ limit of the surfaces in the non-extremal solutions at fixed $r_R$. (near-)Extremal solutions develop a plateau in the $\Delta r$ vs $r_R$ plot, at a value  $\Delta r^-$ that is found to increase with $\alpha$ (Fig. \ref{Fig:10DDeltaPlusDeltaMin}).

In Regime \textbf{II}, empty AdS does not support islands anymore while the extremal solution does. The atoll forms, inducing the change in behavior of $\Delta r$ at larger $r_R$. 
We find that, due to the limitations of the numerical method we use, we cannot reliably obtain island surfaces at large values of $r_R$ when $\alpha >\alpha_c$. The issue is that, while $r_R$ increases freely, $r_L$ remains stuck below the critical anchor. This makes the surfaces stretch more and more, developing a very steep region where $\partial_xr$ is large. With our numerical method, finer grids would be required to reliably characterize these surfaces but we are limited in how much we can increase the resolution. However, we can reach large enough $r_R$ in this regime to infer the main properties of island surfaces.

Again due to numerical limitations, we cannot have a precise quantitative determination of the critical anchor $r_A$. However, we observe that the atoll qualitatively behaves like in the 5d case, with an increasing size $\delta_A$ as rotation increases. This is signaled by the increasing deviation of $\Delta r$ from $r_R-r_h$ at large $r_R$, as rotation increases, which is related to the atoll as:
\begin{equation}
    \Delta r -(r_R-r_h) = r_h-r_A = -\delta_A\,.
\end{equation}

\begin{figure}[]
\centering
\includegraphics[width=0.7\textwidth]{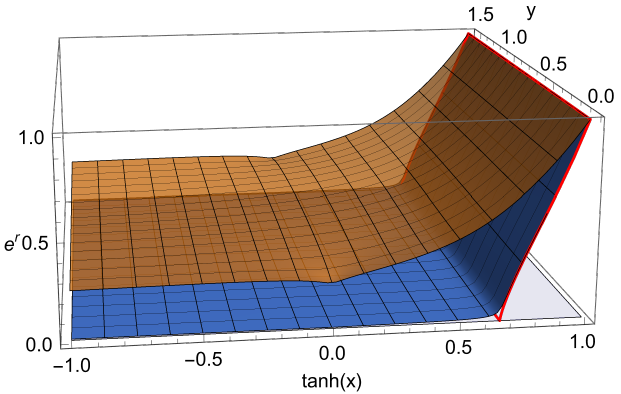}
\caption{Island surfaces in the extremal solution for $\alpha = 5$ (orange) and $\alpha = 7.85\sim\alpha_e$ (blue) and $r_R=0$. The radial coordinate we use is $e^r$ and the ``topological" singularity is located at $e^r=0$, represented by the light blue plane. The bold red line corresponds to the HM surface at $\alpha=7.85$ for a solution close to extremality (with $r_h=-10$). }
\label{Fig:10DExtremalIslands}
\end{figure}

Finally, in regime \textbf{III}, islands cease to exist also for the extremal solution, and $\Delta r^-$ diverges as $\alpha$ approaches $\alpha_e$. We numerically determine $\alpha_e$ by estimating the position of the asymptote in the plot of $\Delta r^-$. As extremality is approached in this regime, the anchor point on the brane $r_L$ for islands that reach the boundary at fixed $r_R$ decreases without bound. Correspondingly $\Delta r$ increases indefinitely. Islands stretch more and more towards the horizon before sharply turning to reach the brane at $r_L \gtrsim r_h $. Exactly at extremality, we expect this sharp turning point to become a cusp, similarly to what happens in the bottom-up setup. This is shown in Fig. \ref{Fig:10DExtremalIslands}, where islands in the extremal black hole geometry are shown for $\alpha = 5<\alpha_e$ (orange) and $\alpha = 7.85 \sim \alpha_e$ (blue) close to becoming singular. Moreover, close to $\alpha_e$ (and in general in regime \textbf{III}), we see that islands and HM surfaces (depicted in red in Fig. \ref{Fig:10DExtremalIslands}) tend to coincide.

\subsubsection{Area difference}
Finally, let us include HM surfaces and compare their area with that of the islands on the initial time slice. A examples of some HM surfaces found in the 10d geometry is shown in Fig. \ref{fig:10DIslands} (right) for $\alpha=2,K=1,r_R = r_h+1$ and different values of $r_h=\{0,-0.5,-1,-1.82\}$, increasing from the innermost to the outermost surface. HM surfaces tend to remain at positive values of $x$. It was found in \cite{Uhlemann:2021nhu} that,  for $\alpha<\alpha_c$, HM surfaces cease to exist beyond a value $r_R = \hat r_R$, slightly larger than $r_R = r^\star$ where island and HM surfaces have the same area. Beyond $\hat r_R$ the surface is pushed to negative values of $x$ and towards $x=-\infty$. The relaxation cease to reach an equilibrium configuration. It was argued in \cite{Uhlemann:2021nhu} that beyond this point, HM surfaces may become saddle points of \eqref{eq:10DInducedAreaHM} that cannot be found by the relaxation method we used, which converges only for minima of the area functional. Furthermore, new minima may appear in this case, extending to negative values of $x$ and with disconnected regions around the five-brane sources. However, these surfaces cannot be globally described by the embedding $x=x(u,y)$.

As we turn on rotation, we find an analogous  behavior for HM surfaces: they cease to exist below $\alpha_c$ beyond a given $\hat r_R$, whose value \textit{decreases with rotation}, as compared to the horizon. This is evident from the fact that HM surfaces are pushed towards $x=0$ by the rotation, as shown in Fig. \ref{fig:10DIslands} where the outermost surfaces, with $r_h =-1.82$, is close to $\hat r_R$. As $\alpha$ increases, also $\hat r_R$ increases and tend to diverge close to $\alpha_c$, beyond which there are no limitations in finding HM surfaces.

\begin{figure}[]
\centering
\includegraphics[width=0.49\textwidth]{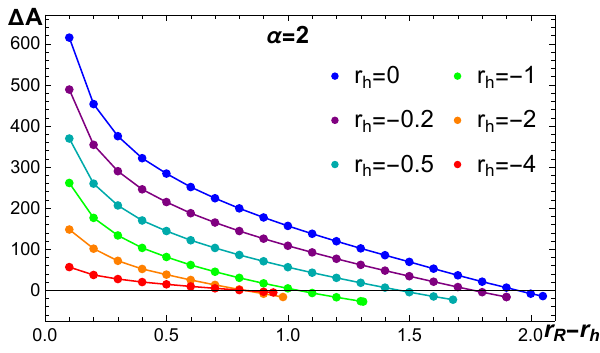}
\ 
\includegraphics[width=0.49\textwidth]{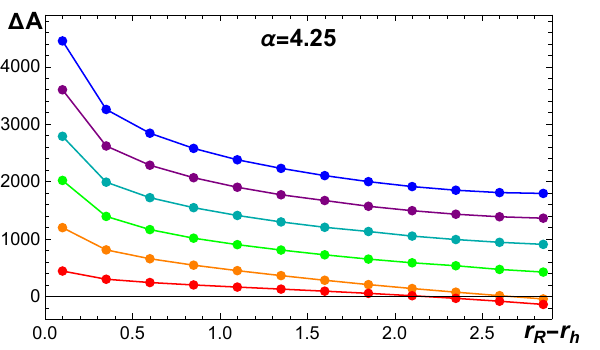}
\caption{$\Delta A$ as a function of $r_R$ for different values of the rotation, this plot is the 10d counterpart of Fig. \ref{Fig:5DAreaDifference}. For $\alpha<\alpha_c$  The end-points of the curves correspond to the values of $\hat r_R$ beyond which HM surfaces cease to exist.  Above $\alpha_c$, HM can always be found but we can only reliably determine $\Delta A$ up to $r_R-r_h \sim 3$.}
\label{Fig:10DAreaDiff}
\end{figure}

Coming back to the area difference, the results are shown in Fig. \ref{Fig:10DAreaDiff} for $\alpha =2$ (regime \textbf{I}) and  $\alpha = 4.25$ (regime \textbf{II}). As in 5d, reliably finding $\Delta A$ becomes increasingly difficult above $\alpha_c$. For $\alpha = 4.25$ we can reach up to $r_R-r_h \sim 3$. Despite these limitations, we find qualitative agreement with the results found in the 5d case. 

Below $\alpha_c$ there always exists a value $r^\star$ at which $\Delta A = 0$, for every value of rotation. The quantity $r^\star$ is pushed towards the horizon as rotation increases. Also, in 10d HM surfaces exist only below $\hat r_R$, corresponding to the end-points of the curves in Fig. \ref{Fig:10DAreaDiff},  which is also pushed towards the horizon with $\Omega$.

Above $\alpha_c$, the behaviour of the curves $\Delta A$ seem to reproduce qualitatively the 5d ones. Curves tend to remain at values $\Delta A >0$, and only for big enough rotation we can find a finite value of $r^\star$. Despite the numerical limitation that forces us to remain below $r_R-r_h\sim 3$, we can see that in this regime curves tend to be lifted and reach an asymptotic finite value at large $r_R$. 

By interpreting this behaviour as the effect of increasing $\alpha$, although we cannot numerically reach higher values of $\alpha$, we expect that the curves enter a new regime when  $\Delta A$ remains above zero for any value of rotation. This would characterize the third regime for which we expect also $\alpha>\alpha_e$.

Notice that as $\alpha$ increases, also $r^\star$ increases and, for  large enough values of $\alpha$, only configurations with enough rotation are characterized by a finite value of $r^\star$.  This should allow us to define a ``Page-tension" $\alpha_p(r_h)$ in analogy to the Page-angle $\mu_p(r_h)$ in 5d, but due to numerical limitations we cannot fully characterize its numerical value.

\section{Conclusions and Outlook}

The study of black hole configurations embedded in a braneworld scenario offers new insights into the structure of extremal surfaces that capture the entanglement of Hawking radiation. 
Our analysis of a rotating cylindrical black hole in AdS has revealed a new critical behavior for extremal surfaces, controlled by a critical angle $\mu_e$. This angle also controls the existence of islands in the extremal solution \eqref{eq:CylindricalBHExtremalMetric}, mirroring the role of the critical angle $\mu_c$ for empty AdS.  

Our results have shown that important properties of the island surfaces' phase space (presence of an atoll, critical behavior at zero temperature) persist for rotating black holes. As expected, and previously pointed out in the literature, the 5d effective braneworld description is consistent with the 10d microscopic derivation only qualitatively, and a quantitative matching of the two setups is prevented by the dependence of the extremal surfaces on details of the internal geometry. 

The fact that two critical values characterizing the system are determined by zero temperature solutions, namely empty AdS and the extremal cylindrical black hole, motivates a further analysis of the properties of extremal solutions in the present context, and the extension of this analysis to other extremal configurations. A study along this direction, together with clarifying the holographic nature of the extremal solution, is left to future work.

The derivation of extremal surfaces in double holography relies on a construction  where the boundary itself contains a black hole, thus the CFT is on a curved background. The state of the CFT on this background is understood to be an Hartle-Hawking like equilibrium state, at finite temperature and angular velocity \cite{Hubeny:2009rc}, and the extremal surfaces we have derived characterize the entanglement phase structure of this state. A similar interpretation is less clear for the extremal solution, which however displays new interesting behaviors, and could be analyzed more extensively in the future.

We point out that, as mentioned in the introduction, the extremal cylindrical black hole does not have a curvature singularity, similarly to the BTZ black hole. It would be interesting to investigate the effect of quantum backreaction of the CFT on the singular metric, in the double holographic construction, along the lines of \cite{Emparan:2020znc,Emparan:2021yon,Bueno:2022log}. 

Given recent progress in studying the embedding in type IIB string theory of  AdS$_4$ vacua \cite{Rovere:2025jks}, like those considered in this work, it will be interesting to see how the entanglement structure we outlined here is modified for other black hole solutions, for example in the presence of scalar fields and charges, in addition to rotation.

\subsection*{Acknowledgements}  

We would like to thank Costas Bachas, Christoph Uhlemann and Achilleas Porfyriadis for interesting discussions. 

\bibliographystyle{JHEP}
\bibliography{braneworlds_biblio}

@article{Lemos:1994xp,
    author = "Lemos, J. P. S.",
    title = "{Cylindrical black hole in general relativity}",
    eprint = "gr-qc/9404041",
    archivePrefix = "arXiv",
    doi = "10.1016/0370-2693(95)00533-Q",
    journal = "Phys. Lett. B",
    volume = "353",
    pages = "46--51",
    year = "1995"
}

@article{Lemos:1995cm,
    author = "Lemos, Jose P. S. and Zanchin, Vilson T.",
    title = "{Rotating charged black string and three-dimensional black holes}",
    eprint = "hep-th/9511188",
    archivePrefix = "arXiv",
    doi = "10.1103/PhysRevD.54.3840",
    journal = "Phys. Rev. D",
    volume = "54",
    pages = "3840--3853",
    year = "1996"
}

@article{Lemos:2000wp,
    author = "Lemos, Jose' P. S.",
    title = "{Supersymmetry of the extreme rotating toroidal black hole}",
    eprint = "hep-th/0011234",
    archivePrefix = "arXiv",
    reportNumber = "CERN-TH-99-402, DF-IST-8-2000",
    doi = "10.1016/S0550-3213(01)00064-5",
    journal = "Nucl. Phys. B",
    volume = "600",
    pages = "272--284",
    year = "2001"
}

@article{Dehghani:2002rr,
    author = "Dehghani, M. H.",
    title = "{Thermodynamics of rotating charged black strings and (A)dS / CFT correspondence}",
    eprint = "hep-th/0205129",
    archivePrefix = "arXiv",
    doi = "10.1103/PhysRevD.66.044006",
    journal = "Phys. Rev. D",
    volume = "66",
    pages = "044006",
    year = "2002"
}

@article{Horowitz:1998ha,
    author = "Horowitz, Gary T. and Myers, Robert C.",
    title = "{The AdS / CFT correspondence and a new positive energy conjecture for general relativity}",
    eprint = "hep-th/9808079",
    archivePrefix = "arXiv",
    reportNumber = "NSF-ITP-98-076, MCGILL-98-13",
    doi = "10.1103/PhysRevD.59.026005",
    journal = "Phys. Rev. D",
    volume = "59",
    pages = "026005",
    year = "1998"
}

@article{Witten:1998zw,
    author = "Witten, Edward",
    editor = "Bergstrom, L. and Lindstrom, U.",
    title = "{Anti-de Sitter space, thermal phase transition, and confinement in gauge theories}",
    eprint = "hep-th/9803131",
    archivePrefix = "arXiv",
    reportNumber = "IASSNS-HEP-98-21",
    doi = "10.4310/ATMP.1998.v2.n3.a3",
    journal = "Adv. Theor. Math. Phys.",
    volume = "2",
    pages = "505--532",
    year = "1998"
}

@article{Surya:2001vj,
    author = "Surya, Sumati and Schleich, Kristin and Witt, Donald M.",
    title = "{Phase transitions for flat AdS black holes}",
    eprint = "hep-th/0101134",
    archivePrefix = "arXiv",
    doi = "10.1103/PhysRevLett.86.5231",
    journal = "Phys. Rev. Lett.",
    volume = "86",
    pages = "5231--5234",
    year = "2001"
}

@article{Marolf:2013ioa,
    author = "Marolf, Donald and Rangamani, Mukund and Wiseman, Toby",
    title = "{Holographic thermal field theory on curved spacetimes}",
    eprint = "1312.0612",
    archivePrefix = "arXiv",
    primaryClass = "hep-th",
    reportNumber = "DCPT-13-51",
    doi = "10.1088/0264-9381/31/6/063001",
    journal = "Class. Quant. Grav.",
    volume = "31",
    pages = "063001",
    year = "2014"
}

@article{Randall:1999vf,
    author = "Randall, Lisa and Sundrum, Raman",
    title = "{An Alternative to compactification}",
    eprint = "hep-th/9906064",
    archivePrefix = "arXiv",
    reportNumber = "MIT-CTP-2874, PUPT-1867, BUHEP-99-13",
    doi = "10.1103/PhysRevLett.83.4690",
    journal = "Phys. Rev. Lett.",
    volume = "83",
    pages = "4690--4693",
    year = "1999"
}

@article{Karch:2000ct,
    author = "Karch, Andreas and Randall, Lisa",
    editor = "Duff, Michael J. and Liu, J. T. and Lu, J.",
    title = "{Locally localized gravity}",
    eprint = "hep-th/0011156",
    archivePrefix = "arXiv",
    reportNumber = "MIT-CTP-3099",
    doi = "10.1088/1126-6708/2001/05/008",
    journal = "JHEP",
    volume = "05",
    pages = "008",
    year = "2001"
}

@article{deHaro:2000wj,
    author = "de Haro, Sebastian and Skenderis, Kostas and Solodukhin, Sergey N.",
    editor = "Semikhatov, A. and Vasilev, M. and Zaikin, V.",
    title = "{Gravity in warped compactifications and the holographic stress tensor}",
    eprint = "hep-th/0011230",
    archivePrefix = "arXiv",
    reportNumber = "SPIN-2000-30, ITP-UU-00-32, PUTP-1970",
    doi = "10.1088/0264-9381/18/16/307",
    journal = "Class. Quant. Grav.",
    volume = "18",
    pages = "3171--3180",
    year = "2001"
}

@article{Karch:2000gx,
    author = "Karch, Andreas and Randall, Lisa",
    title = "{Open and closed string interpretation of SUSY CFT's on branes with boundaries}",
    eprint = "hep-th/0105132",
    archivePrefix = "arXiv",
    reportNumber = "MIT-CTP-3146",
    doi = "10.1088/1126-6708/2001/06/063",
    journal = "JHEP",
    volume = "06",
    pages = "063",
    year = "2001"
}

@article{Takayanagi:2011zk,
    author = "Takayanagi, Tadashi",
    title = "{Holographic Dual of BCFT}",
    eprint = "1105.5165",
    archivePrefix = "arXiv",
    primaryClass = "hep-th",
    reportNumber = "IPMU11-0091",
    doi = "10.1103/PhysRevLett.107.101602",
    journal = "Phys. Rev. Lett.",
    volume = "107",
    pages = "101602",
    year = "2011"
}

@article{Fujita:2011fp,
    author = "Fujita, Mitsutoshi and Takayanagi, Tadashi and Tonni, Erik",
    title = "{Aspects of AdS/BCFT}",
    eprint = "1108.5152",
    archivePrefix = "arXiv",
    primaryClass = "hep-th",
    reportNumber = "IPMU-11-0136, MIT-CTP-4289",
    doi = "10.1007/JHEP11(2011)043",
    journal = "JHEP",
    volume = "11",
    pages = "043",
    year = "2011"
}

@article{Omiya:2021olc,
    author = "Omiya, Hidetoshi and Wei, Zixia",
    title = "{Causal structures and nonlocality in double holography}",
    eprint = "2107.01219",
    archivePrefix = "arXiv",
    primaryClass = "hep-th",
    reportNumber = "YITP-21-51",
    doi = "10.1007/JHEP07(2022)128",
    journal = "JHEP",
    volume = "07",
    pages = "128",
    year = "2022"
}

@article{Neuenfeld:2023svs,
    author = "Neuenfeld, Dominik and Srivastava, Manu",
    title = "{On the causality paradox and the Karch-Randall braneworld as an EFT}",
    eprint = "2307.10392",
    archivePrefix = "arXiv",
    primaryClass = "hep-th",
    reportNumber = "MIT-CTP/5586",
    doi = "10.1007/JHEP10(2023)164",
    journal = "JHEP",
    volume = "10",
    pages = "164",
    year = "2023"
}

@article{Mori:2023swn,
    author = "Mori, Takato and Yoshida, Beni",
    title = "{Exploring causality in braneworld/cutoff holography via holographic scattering}",
    eprint = "2308.00739",
    archivePrefix = "arXiv",
    primaryClass = "hep-th",
    reportNumber = "YITP-23-94",
    doi = "10.1007/JHEP10(2023)104",
    journal = "JHEP",
    volume = "10",
    pages = "104",
    year = "2023"
}

@article{Geng:2025yys,
    author = "Geng, Hao and Randall, Lisa",
    title = "{Holography and Causality in the Karch-Randall Braneworld}",
    eprint = "2504.21856",
    archivePrefix = "arXiv",
    primaryClass = "hep-th",
    month = "4",
    year = "2025"
}

@article{Porrati:2003sa,
    author = "Porrati, M.",
    title = "{Higgs phenomenon for the graviton in ADS space}",
    eprint = "hep-th/0306253",
    archivePrefix = "arXiv",
    doi = "10.1142/S0217732303011745",
    journal = "Mod. Phys. Lett. A",
    volume = "18",
    pages = "1793--1802",
    year = "2003"
}

@article{Porrati:2001gx,
    author = "Porrati, Massimo",
    title = "{Mass and gauge invariance 4. Holography for the Karch-Randall model}",
    eprint = "hep-th/0109017",
    archivePrefix = "arXiv",
    reportNumber = "NYU-TH-01-08-02",
    doi = "10.1103/PhysRevD.65.044015",
    journal = "Phys. Rev. D",
    volume = "65",
    pages = "044015",
    year = "2002"
}

@article{Porrati:2002dt,
    author = "Porrati, Massimo and Starinets, Andrei",
    title = "{On the graviton selfenergy in AdS(4)}",
    eprint = "hep-th/0201261",
    archivePrefix = "arXiv",
    reportNumber = "NYU-TH-02-01-01, CU-TP-1049",
    doi = "10.1016/S0370-2693(02)01490-9",
    journal = "Phys. Lett. B",
    volume = "532",
    pages = "48--54",
    year = "2002"
}

@article{Aharony:2006hz,
    author = "Aharony, Ofer and Clark, Adam B. and Karch, Andreas",
    title = "{The CFT/AdS correspondence, massive gravitons and a connectivity index conjecture}",
    eprint = "hep-th/0608089",
    archivePrefix = "arXiv",
    reportNumber = "WIS-09-06-AUG-DPP",
    doi = "10.1103/PhysRevD.74.086006",
    journal = "Phys. Rev. D",
    volume = "74",
    pages = "086006",
    year = "2006"
}

@article{Mathur:2014dia,
    author = "Mathur, Samir D.",
    title = "{What is the dual of two entangled CFTs?}",
    eprint = "1402.6378",
    archivePrefix = "arXiv",
    primaryClass = "hep-th",
    month = "2",
    year = "2014"
}

@article{Rocha:2008fe,
    author = "Rocha, Jorge V.",
    title = "{Evaporation of large black holes in AdS: Coupling to the evaporon}",
    eprint = "0804.0055",
    archivePrefix = "arXiv",
    primaryClass = "hep-th",
    doi = "10.1088/1126-6708/2008/08/075",
    journal = "JHEP",
    volume = "08",
    pages = "075",
    year = "2008"
}

@article{Ryu:2006bv,
    author = "Ryu, Shinsei and Takayanagi, Tadashi",
    title = "{Holographic derivation of entanglement entropy from AdS/CFT}",
    eprint = "hep-th/0603001",
    archivePrefix = "arXiv",
    reportNumber = "NSF-KITP-06-11",
    doi = "10.1103/PhysRevLett.96.181602",
    journal = "Phys. Rev. Lett.",
    volume = "96",
    pages = "181602",
    year = "2006"
}

@article{Hubeny:2007xt,
    author = "Hubeny, Veronika E. and Rangamani, Mukund and Takayanagi, Tadashi",
    title = "{A Covariant holographic entanglement entropy proposal}",
    eprint = "0705.0016",
    archivePrefix = "arXiv",
    primaryClass = "hep-th",
    reportNumber = "DCPT-07-13, KUNS-2069",
    doi = "10.1088/1126-6708/2007/07/062",
    journal = "JHEP",
    volume = "07",
    pages = "062",
    year = "2007"
}

@article{Hartman:2013qma,
    author = "Hartman, Thomas and Maldacena, Juan",
    title = "{Time Evolution of Entanglement Entropy from Black Hole Interiors}",
    eprint = "1303.1080",
    archivePrefix = "arXiv",
    primaryClass = "hep-th",
    doi = "10.1007/JHEP05(2013)014",
    journal = "JHEP",
    volume = "05",
    pages = "014",
    year = "2013"
}

@article{Engelhardt:2014gca,
    author = "Engelhardt, Netta and Wall, Aron C.",
    title = "{Quantum Extremal Surfaces: Holographic Entanglement Entropy beyond the Classical Regime}",
    eprint = "1408.3203",
    archivePrefix = "arXiv",
    primaryClass = "hep-th",
    doi = "10.1007/JHEP01(2015)073",
    journal = "JHEP",
    volume = "01",
    pages = "073",
    year = "2015"
}

@article{Almheiri:2019psf,
    author = "Almheiri, Ahmed and Engelhardt, Netta and Marolf, Donald and Maxfield, Henry",
    title = "{The entropy of bulk quantum fields and the entanglement wedge of an evaporating black hole}",
    eprint = "1905.08762",
    archivePrefix = "arXiv",
    primaryClass = "hep-th",
    doi = "10.1007/JHEP12(2019)063",
    journal = "JHEP",
    volume = "12",
    pages = "063",
    year = "2019"
}

@article{Penington:2019npb,
    author = "Penington, Geoffrey",
    title = "{Entanglement Wedge Reconstruction and the Information Paradox}",
    eprint = "1905.08255",
    archivePrefix = "arXiv",
    primaryClass = "hep-th",
    doi = "10.1007/JHEP09(2020)002",
    journal = "JHEP",
    volume = "09",
    pages = "002",
    year = "2020"
}

@article{Almheiri:2019hni,
    author = "Almheiri, Ahmed and Mahajan, Raghu and Maldacena, Juan and Zhao, Ying",
    title = "{The Page curve of Hawking radiation from semiclassical geometry}",
    eprint = "1908.10996",
    archivePrefix = "arXiv",
    primaryClass = "hep-th",
    doi = "10.1007/JHEP03(2020)149",
    journal = "JHEP",
    volume = "03",
    pages = "149",
    year = "2020"
}

@article{Almheiri:2019yqk,
    author = "Almheiri, Ahmed and Mahajan, Raghu and Maldacena, Juan",
    title = "{Islands outside the horizon}",
    eprint = "1910.11077",
    archivePrefix = "arXiv",
    primaryClass = "hep-th",
    month = "10",
    year = "2019"
}

@article{Almheiri:2019psy,
    author = "Almheiri, Ahmed and Mahajan, Raghu and Santos, Jorge E.",
    title = "{Entanglement islands in higher dimensions}",
    eprint = "1911.09666",
    archivePrefix = "arXiv",
    primaryClass = "hep-th",
    doi = "10.21468/SciPostPhys.9.1.001",
    journal = "SciPost Phys.",
    volume = "9",
    number = "1",
    pages = "001",
    year = "2020"
}

@article{Chen:2020uac,
    author = "Chen, Hong Zhe and Myers, Robert C. and Neuenfeld, Dominik and Reyes, Ignacio A. and Sandor, Joshua",
    title = "{Quantum Extremal Islands Made Easy, Part I: Entanglement on the Brane}",
    eprint = "2006.04851",
    archivePrefix = "arXiv",
    primaryClass = "hep-th",
    doi = "10.1007/JHEP10(2020)166",
    journal = "JHEP",
    volume = "10",
    pages = "166",
    year = "2020"
}

@article{Geng:2025byh,
    author = "Geng, Hao",
    title = "{Making the Case for Massive Islands}",
    eprint = "2509.22775",
    archivePrefix = "arXiv",
    primaryClass = "hep-th",
    month = "9",
    year = "2025"
}

@article{Geng:2020qvw,
    author = "Geng, Hao and Karch, Andreas",
    title = "{Massive islands}",
    eprint = "2006.02438",
    archivePrefix = "arXiv",
    primaryClass = "hep-th",
    doi = "10.1007/JHEP09(2020)121",
    journal = "JHEP",
    volume = "09",
    pages = "121",
    year = "2020"
}

@article{Geng:2020fxl,
    author = "Geng, Hao and Karch, Andreas and Perez-Pardavila, Carlos and Raju, Suvrat and Randall, Lisa and Riojas, Marcos and Shashi, Sanjit",
    title = "{Information Transfer with a Gravitating Bath}",
    eprint = "2012.04671",
    archivePrefix = "arXiv",
    primaryClass = "hep-th",
    doi = "10.21468/SciPostPhys.10.5.103",
    journal = "SciPost Phys.",
    volume = "10",
    number = "5",
    pages = "103",
    year = "2021"
}

@article{Chen:2020hmv,
    author = "Chen, Hong Zhe and Myers, Robert C. and Neuenfeld, Dominik and Reyes, Ignacio A. and Sandor, Joshua",
    title = "{Quantum Extremal Islands Made Easy, Part II: Black Holes on the Brane}",
    eprint = "2010.00018",
    archivePrefix = "arXiv",
    primaryClass = "hep-th",
    doi = "10.1007/JHEP12(2020)025",
    journal = "JHEP",
    volume = "12",
    pages = "025",
    year = "2020"
}

@article{Geng:2021mic,
    author = "Geng, Hao and Karch, Andreas and Perez-Pardavila, Carlos and Raju, Suvrat and Randall, Lisa and Riojas, Marcos and Shashi, Sanjit",
    title = "{Entanglement phase structure of a holographic BCFT in a black hole background}",
    eprint = "2112.09132",
    archivePrefix = "arXiv",
    primaryClass = "hep-th",
    reportNumber = "UTTG-27-2021",
    doi = "10.1007/JHEP05(2022)153",
    journal = "JHEP",
    volume = "05",
    pages = "153",
    year = "2022"
}

@article{Karch:2023ekf,
    author = "Karch, Andreas and Perez-Pardavila, Carlos and Riojas, Marcos and Youssef, Merna",
    title = "{Subregion entropy for the doubly-holographic global black string}",
    eprint = "2303.09571",
    archivePrefix = "arXiv",
    primaryClass = "hep-th",
    reportNumber = "UTWI-08-2023",
    doi = "10.1007/JHEP05(2023)195",
    journal = "JHEP",
    volume = "05",
    pages = "195",
    year = "2023"
}

@article{Geng:2023qwm,
    author = "Geng, Hao",
    title = "{Revisiting Recent Progress in the Karch-Randall Braneworld}",
    eprint = "2306.15671",
    archivePrefix = "arXiv",
    primaryClass = "hep-th",
    month = "6",
    year = "2023"
}

@article{Geng:2021hlu,
    author = "Geng, Hao and Karch, Andreas and Perez-Pardavila, Carlos and Raju, Suvrat and Randall, Lisa and Riojas, Marcos and Shashi, Sanjit",
    title = "{Inconsistency of islands in theories with long-range gravity}",
    eprint = "2107.03390",
    archivePrefix = "arXiv",
    primaryClass = "hep-th",
    doi = "10.1007/JHEP01(2022)182",
    journal = "JHEP",
    volume = "01",
    pages = "182",
    year = "2022"
}

@article{DHoker:2007zhm,
    author = "D'Hoker, Eric and Estes, John and Gutperle, Michael",
    title = "{Exact half-BPS Type IIB interface solutions. I. Local solution and supersymmetric Janus}",
    eprint = "0705.0022",
    archivePrefix = "arXiv",
    primaryClass = "hep-th",
    reportNumber = "UCLA-07-TEP-09",
    doi = "10.1088/1126-6708/2007/06/021",
    journal = "JHEP",
    volume = "06",
    pages = "021",
    year = "2007"
}

@article{DHoker:2007hhe,
    author = "D'Hoker, Eric and Estes, John and Gutperle, Michael",
    title = "{Exact half-BPS Type IIB interface solutions. II. Flux solutions and multi-Janus}",
    eprint = "0705.0024",
    archivePrefix = "arXiv",
    primaryClass = "hep-th",
    reportNumber = "UCLA-07-TEP-10",
    doi = "10.1088/1126-6708/2007/06/022",
    journal = "JHEP",
    volume = "06",
    pages = "022",
    year = "2007"
}

@article{Aharony:2011yc,
    author = "Aharony, Ofer and Berdichevsky, Leon and Berkooz, Micha and Shamir, Itamar",
    title = "{Near-horizon solutions for D3-branes ending on 5-branes}",
    eprint = "1106.1870",
    archivePrefix = "arXiv",
    primaryClass = "hep-th",
    reportNumber = "WIS-5-11-MAY-DPPA",
    doi = "10.1103/PhysRevD.84.126003",
    journal = "Phys. Rev. D",
    volume = "84",
    pages = "126003",
    year = "2011"
}

@article{Assel:2011xz,
    author = "Assel, Benjamin and Bachas, Costas and Estes, John and Gomis, Jaume",
    title = "{Holographic Duals of D=3 N=4 Superconformal Field Theories}",
    eprint = "1106.4253",
    archivePrefix = "arXiv",
    primaryClass = "hep-th",
    doi = "10.1007/JHEP08(2011)087",
    journal = "JHEP",
    volume = "08",
    pages = "087",
    year = "2011"
}

@article{Bachas:2018zmb,
    author = "Bachas, Constantin and Lavdas, Ioannis",
    title = "{Massive Anti-de Sitter Gravity from String Theory}",
    eprint = "1807.00591",
    archivePrefix = "arXiv",
    primaryClass = "hep-th",
    doi = "10.1007/JHEP11(2018)003",
    journal = "JHEP",
    volume = "11",
    pages = "003",
    year = "2018"
}

@article{Karch:2022rvr,
    author = "Karch, Andreas and Sun, Haoyu and Uhlemann, Christoph F.",
    title = "{Double holography in string theory}",
    eprint = "2206.11292",
    archivePrefix = "arXiv",
    primaryClass = "hep-th",
    reportNumber = "LCTP-22-08",
    doi = "10.1007/JHEP10(2022)012",
    journal = "JHEP",
    volume = "10",
    pages = "012",
    year = "2022"
}

@article{Uhlemann:2021nhu,
    author = "Uhlemann, Christoph F.",
    title = "{Islands and Page curves in 4d from Type IIB}",
    eprint = "2105.00008",
    archivePrefix = "arXiv",
    primaryClass = "hep-th",
    reportNumber = "LCTP-21-09",
    doi = "10.1007/JHEP08(2021)104",
    journal = "JHEP",
    volume = "08",
    pages = "104",
    year = "2021"
}

@article{Demulder:2022aij,
    author = "Demulder, Saskia and Gnecchi, Alessandra and Lavdas, Ioannis and Lust, Dieter",
    title = "{Islands and light gravitons in type IIB string theory}",
    eprint = "2204.03669",
    archivePrefix = "arXiv",
    primaryClass = "hep-th",
    reportNumber = "LMU-ASC 14/22, MPP-2022-40",
    doi = "10.1007/JHEP02(2023)016",
    journal = "JHEP",
    volume = "02",
    pages = "016",
    year = "2023"
}

@article{Deddo:2023oxn,
    author = "Deddo, Evan and Pando Zayas, Leopoldo A. and Uhlemann, Christoph F.",
    title = "{Binary AdS black holes coupled to a bath in Type IIB}",
    eprint = "2401.00511",
    archivePrefix = "arXiv",
    primaryClass = "hep-th",
    reportNumber = "LCTP-23-19",
    doi = "10.1007/JHEP05(2024)120",
    journal = "JHEP",
    volume = "05",
    pages = "120",
    year = "2024"
}

@article{Nian:2019buz,
    author = "Nian, Jun",
    title = "{Kerr black hole evaporation and Page curve}",
    eprint = "1912.13474",
    archivePrefix = "arXiv",
    primaryClass = "hep-th",
    reportNumber = "LCTP-19-38",
    doi = "10.1142/S0218271824500305",
    journal = "Int. J. Mod. Phys. D",
    volume = "33",
    number = "07n08",
    pages = "2450030",
    year = "2024"
}

@article{Nian:2023xmr,
    author = "Nian, Jun",
    title = "{Hawking Radiation, Entanglement Entropy, and Information Paradox of Kerr Black Holes}",
    eprint = "2312.14287",
    archivePrefix = "arXiv",
    primaryClass = "hep-th",
    month = "12",
    year = "2023"
}

@article{Yu:2021rfg,
    author = "Yu, Ming-Hui and Lu, Cheng-Yuan and Ge, Xian-Hui and Sin, Sang-Jin",
    title = "{Island, Page curve, and superradiance of rotating BTZ black holes}",
    eprint = "2112.14361",
    archivePrefix = "arXiv",
    primaryClass = "hep-th",
    doi = "10.1103/PhysRevD.105.066009",
    journal = "Phys. Rev. D",
    volume = "105",
    number = "6",
    pages = "066009",
    year = "2022"
}

@article{Wang:2024itz,
    author = "Wang, Liqiang and Li, Ran",
    title = "{Entanglement islands and the Page curve of Hawking radiation for rotating Kerr black holes}",
    eprint = "2406.13949",
    archivePrefix = "arXiv",
    primaryClass = "hep-th",
    doi = "10.1103/PhysRevD.110.066012",
    journal = "Phys. Rev. D",
    volume = "110",
    number = "6",
    pages = "066012",
    year = "2024"
}

@article{Yu:2025euq,
    author = "Yu, Ming-Hui and Ge, Xian-Hui",
    title = "{Islands in Kerr-Newman Black Holes}",
    eprint = "2510.24006",
    archivePrefix = "arXiv",
    primaryClass = "hep-th",
    month = "10",
    year = "2025"
}

@article{Gregory:1993vy,
    author = "Gregory, R. and Laflamme, R.",
    title = "{Black strings and p-branes are unstable}",
    eprint = "hep-th/9301052",
    archivePrefix = "arXiv",
    doi = "10.1103/PhysRevLett.70.2837",
    journal = "Phys. Rev. Lett.",
    volume = "70",
    pages = "2837--2840",
    year = "1993"
}

@article{Gregory:1994bj,
    author = "Gregory, Ruth and Laflamme, Raymond",
    title = "{The Instability of charged black strings and p-branes}",
    eprint = "hep-th/9404071",
    archivePrefix = "arXiv",
    reportNumber = "DAMTP-R-94-7, LA-UR-93-4473",
    doi = "10.1016/0550-3213(94)90206-2",
    journal = "Nucl. Phys. B",
    volume = "428",
    pages = "399--434",
    year = "1994"
}

@article{Gubser:2000mm,
    author = "Gubser, Steven S. and Mitra, Indrajit",
    title = "{The Evolution of unstable black holes in anti-de Sitter space}",
    eprint = "hep-th/0011127",
    archivePrefix = "arXiv",
    reportNumber = "PUPT-1966",
    doi = "10.1088/1126-6708/2001/08/018",
    journal = "JHEP",
    volume = "08",
    pages = "018",
    year = "2001"
}

@article{Chamblin:2004vr,
    author = "Chamblin, A. and Karch, A.",
    title = "{Hawking and Page on the brane}",
    eprint = "hep-th/0412017",
    archivePrefix = "arXiv",
    reportNumber = "UW-PT-04-24",
    doi = "10.1103/PhysRevD.72.066011",
    journal = "Phys. Rev. D",
    volume = "72",
    pages = "066011",
    year = "2005"
}

@article{Chen:2008vh,
    author = "Chen, Si and Schleich, Kristin and Witt, Donald M.",
    title = "{Thermodynamics and Stability of Flat Anti-de Sitter Black Strings}",
    eprint = "0809.1367",
    archivePrefix = "arXiv",
    primaryClass = "hep-th",
    doi = "10.1103/PhysRevD.78.126001",
    journal = "Phys. Rev. D",
    volume = "78",
    pages = "126001",
    year = "2008"
}

@article{Hirayama:2001bi,
    author = "Hirayama, Takayuki and Kang, Gungwon",
    title = "{Stable black strings in anti-de Sitter space}",
    eprint = "hep-th/0104213",
    archivePrefix = "arXiv",
    reportNumber = "KEK-TH-762",
    doi = "10.1103/PhysRevD.64.064010",
    journal = "Phys. Rev. D",
    volume = "64",
    pages = "064010",
    year = "2001"
}

@article{Marolf:2019wkz,
    author = "Marolf, Donald and Santos, Jorge E.",
    title = "{Phases of Holographic Hawking Radiation on spatially compact spacetimes}",
    eprint = "1906.07681",
    archivePrefix = "arXiv",
    primaryClass = "hep-th",
    doi = "10.1007/JHEP10(2019)250",
    journal = "JHEP",
    volume = "10",
    pages = "250",
    year = "2019"
}

@article{Ghosh:2021axl,
    author = "Ghosh, Kausik and Krishnan, Chethan",
    title = "{Dirichlet baths and the not-so-fine-grained Page curve}",
    eprint = "2103.17253",
    archivePrefix = "arXiv",
    primaryClass = "hep-th",
    doi = "10.1007/JHEP08(2021)119",
    journal = "JHEP",
    volume = "08",
    pages = "119",
    year = "2021"
}

@article{Harvey:2025ttz,
    author = "Harvey, William and Jensen, Kristan and Uzu, Takahiro",
    title = "{Comparing top-down and bottom-up holographic defects and boundaries}",
    eprint = "2504.13244",
    archivePrefix = "arXiv",
    primaryClass = "hep-th",
    month = "4",
    year = "2025"
}

@article{Bhattacharya:2025tdn,
    author = "Bhattacharya, Dyuman and Hennigar, Robie A. and Mann, Robert B. and Zhang, Ming",
    title = "{Charged rotating quantum black holes}",
    eprint = "2506.19941",
    archivePrefix = "arXiv",
    primaryClass = "hep-th",
    month = "6",
    year = "2025"
}

@article{Hubeny:2009rc,
    author = "Hubeny, Veronika E. and Marolf, Donald and Rangamani, Mukund",
    title = "{Hawking radiation from AdS black holes}",
    eprint = "0911.4144",
    archivePrefix = "arXiv",
    primaryClass = "hep-th",
    reportNumber = "DCPT-09-77, DCPT-09/77, NSF-KITP-09-204",
    doi = "10.1088/0264-9381/27/9/095018",
    journal = "Class. Quant. Grav.",
    volume = "27",
    pages = "095018",
    year = "2010"
}

@article{Hubeny:2009ru,
    author = "Hubeny, Veronika E. and Marolf, Donald and Rangamani, Mukund",
    title = "{Hawking radiation in large N strongly-coupled field theories}",
    eprint = "0908.2270",
    archivePrefix = "arXiv",
    primaryClass = "hep-th",
    reportNumber = "DCPT-09-57, NSF-KITP-09-164",
    doi = "10.1088/0264-9381/27/9/095015",
    journal = "Class. Quant. Grav.",
    volume = "27",
    pages = "095015",
    year = "2010"
}

@article{Almheiri:2020cfm,
    author = "Almheiri, Ahmed and Hartman, Thomas and Maldacena, Juan and Shaghoulian, Edgar and Tajdini, Amirhossein",
    title = "{The entropy of Hawking radiation}",
    eprint = "2006.06872",
    archivePrefix = "arXiv",
    primaryClass = "hep-th",
    doi = "10.1103/RevModPhys.93.035002",
    journal = "Rev. Mod. Phys.",
    volume = "93",
    number = "3",
    pages = "035002",
    year = "2021"
}

@article{Faulkner:2013ana,
    author = "Faulkner, Thomas and Lewkowycz, Aitor and Maldacena, Juan",
    title = "{Quantum corrections to holographic entanglement entropy}",
    eprint = "1307.2892",
    archivePrefix = "arXiv",
    primaryClass = "hep-th",
    doi = "10.1007/JHEP11(2013)074",
    journal = "JHEP",
    volume = "11",
    pages = "074",
    year = "2013"
}

@article{Hawking:1982dh,
    author = "Hawking, S. W. and Page, Don N.",
    title = "{Thermodynamics of Black Holes in anti-De Sitter Space}",
    reportNumber = "PRINT-83-0019 (CAMBRIDGE)",
    doi = "10.1007/BF01208266",
    journal = "Commun. Math. Phys.",
    volume = "87",
    pages = "577",
    year = "1983"
}

@article{Hawking:1998kw,
    author = "Hawking, S. W. and Hunter, C. J. and Taylor, Marika",
    title = "{Rotation and the AdS / CFT correspondence}",
    eprint = "hep-th/9811056",
    archivePrefix = "arXiv",
    doi = "10.1103/PhysRevD.59.064005",
    journal = "Phys. Rev. D",
    volume = "59",
    pages = "064005",
    year = "1999"
}

@article{He:2025due,
    author = "He, Dongming and Uhlemann, Christoph F.",
    title = "{One-point functions for doubly-holographic BCFTs and backreacting defects}",
    eprint = "2501.07630",
    archivePrefix = "arXiv",
    primaryClass = "hep-th",
    doi = "10.1007/jhep05(2025)227",
    journal = "JHEP",
    volume = "25",
    number = "5",
    pages = "227",
    year = "2025"
}

@article{Rozali:2019day,
    author = "Rozali, Moshe and Sully, James and Van Raamsdonk, Mark and Waddell, Christopher and Wakeham, David",
    title = "{Information radiation in BCFT models of black holes}",
    eprint = "1910.12836",
    archivePrefix = "arXiv",
    primaryClass = "hep-th",
    doi = "10.1007/JHEP05(2020)004",
    journal = "JHEP",
    volume = "05",
    pages = "004",
    year = "2020"
}

@article{Emparan:2020znc,
    author = "Emparan, Roberto and Frassino, Antonia Micol and Way, Benson",
    title = "{Quantum BTZ black hole}",
    eprint = "2007.15999",
    archivePrefix = "arXiv",
    primaryClass = "hep-th",
    doi = "10.1007/JHEP11(2020)137",
    journal = "JHEP",
    volume = "11",
    pages = "137",
    year = "2020"
}

@article{Emparan:2021yon,
    author = "Emparan, Roberto and Toma{\v{s}}evi{\'c}, Marija",
    title = "{Quantum backreaction on chronology horizons}",
    eprint = "2109.03611",
    archivePrefix = "arXiv",
    primaryClass = "hep-th",
    doi = "10.1007/JHEP02(2022)182",
    journal = "JHEP",
    volume = "02",
    pages = "182",
    year = "2022"
}

@article{Bueno:2022log,
    author = "Bueno, Pablo and Emparan, Roberto and Llorens, Quim",
    title = "{Higher-curvature gravities from braneworlds and the holographic c-theorem}",
    eprint = "2204.13421",
    archivePrefix = "arXiv",
    primaryClass = "hep-th",
    reportNumber = "CERN-TH-2022-076",
    doi = "10.1103/PhysRevD.106.044012",
    journal = "Phys. Rev. D",
    volume = "106",
    number = "4",
    pages = "044012",
    year = "2022"
}

@misc{Mathematica,
  author = "{Wolfram Research{,} Inc.}",
  title = {Mathematica, {V}ersion 14.0},
  url = {https://www.wolfram.com/mathematica},
  note = {Champaign, IL, 2024}
}

@article{Raamsdonk:2020tin,
    author = "Raamsdonk, Mark Van and Waddell, Chris",
    title = "{Holographic and localization calculations of boundary F for $ \mathcal{N} $ = 4 SUSY Yang-Mills theory}",
    eprint = "2010.14520",
    archivePrefix = "arXiv",
    primaryClass = "hep-th",
    doi = "10.1007/JHEP02(2021)222",
    journal = "JHEP",
    volume = "02",
    pages = "222",
    year = "2021"
}

@article{VanRaamsdonk:2021duo,
    author = "Van Raamsdonk, Mark and Waddell, Chris",
    title = "{Finding AdS$^{5}${\texttimes} S$^{5}$ in 2+1 dimensional SCFT physics}",
    eprint = "2109.04479",
    archivePrefix = "arXiv",
    primaryClass = "hep-th",
    doi = "10.1007/JHEP11(2021)145",
    journal = "JHEP",
    volume = "11",
    pages = "145",
    year = "2021"
}

@article{DeLuca:2023kjj,
    author = "De Luca, G. Bruno and De Ponti, Nicol{\`o} and Mondino, Andrea and Tomasiello, Alessandro",
    title = "{Harmonic functions and gravity localization}",
    eprint = "2306.05456",
    archivePrefix = "arXiv",
    primaryClass = "hep-th",
    doi = "10.1007/JHEP09(2023)127",
    journal = "JHEP",
    volume = "09",
    pages = "127",
    year = "2023"
}

@article{Huertas:2023syg,
    author = "Huertas, Jes{\'u}s and Uranga, Angel M.",
    title = "{Aspects of dynamical cobordism in AdS/CFT}",
    eprint = "2306.07335",
    archivePrefix = "arXiv",
    primaryClass = "hep-th",
    doi = "10.1007/JHEP08(2023)140",
    journal = "JHEP",
    volume = "08",
    pages = "140",
    year = "2023"
}

@article{Chaney:2024bgx,
    author = "Chaney, Andrea and Uhlemann, Christoph F.",
    title = "{BMN-like sectors in 4d $ \mathcal{N} $ = 4 SYM with boundaries and interfaces}",
    eprint = "2408.12651",
    archivePrefix = "arXiv",
    primaryClass = "hep-th",
    doi = "10.1007/JHEP01(2025)024",
    journal = "JHEP",
    volume = "01",
    pages = "024",
    year = "2025"
}

@article{Anastasi:2025puv,
    author = "Anastasi, Edoardo and Angius, Roberta and Huertas, Jes{\'u}s and Uranga, Angel M. and Wang, Chuying",
    title = "{Relative quantum gravity: localized gravity and the swampland}",
    eprint = "2501.03310",
    archivePrefix = "arXiv",
    primaryClass = "hep-th",
    doi = "10.1007/JHEP08(2025)107",
    journal = "JHEP",
    volume = "08",
    pages = "107",
    year = "2025"
}

@article{Rovere:2025jks,
    author = "Rovere, Davide and Sterckx, Colin",
    title = "{How to uplift non-maximal gauged supergravities}",
    eprint = "2510.24850",
    archivePrefix = "arXiv",
    primaryClass = "hep-th",
    month = "10",
    year = "2025"
}

@article{Geng:2025gns,
    author = "Geng, Hao and Huertas, Jes{\'u}s and Karch, Andreas and Randall, Lisa and Thomas, Dawson",
    title = "{Wet Hair: Global Symmetries in Entanglement Islands}",
    eprint = "2512.11025",
    archivePrefix = "arXiv",
    primaryClass = "hep-th",
    month = "12",
    year = "2025"
}

@article{Awad:2002cz,
    author = "Awad, Adel M.",
    title = "{Higher dimensional charged rotating solutions in (A)dS space-times}",
    eprint = "hep-th/0209238",
    archivePrefix = "arXiv",
    reportNumber = "UCTP-105-02",
    doi = "10.1088/0264-9381/20/13/327",
    journal = "Class. Quant. Grav.",
    volume = "20",
    pages = "2827--2834",
    year = "2003"
}

\end{document}